\documentclass[twocolumn]{aastex631}

\usepackage{amsmath}

\received{2026.07.09.}
\revised{2026.09.10.}
\accepted{2026.09.22.}

\submitjournal{PASP}

\begin{document}

\title{
Limits on eccentricity excitation by tidally enhanced mass loss in Wolf–Rayet binaries
}

\correspondingauthor{Viktória Fröhlich}
\email{frohlich.viktoria@csfk.org}

\author[0000-0003-3780-7185]{Viktória Fröhlich}
\affiliation{HUN-REN CSFK Konkoly Observatory, MTA Centre of Excellence, Konkoly Thege M. \'ut 15-17, Budapest, 1121, Hungary}
\affiliation{ELTE Eötvös Loránd University, Institute of Physics and Astronomy, P\'azm\'any P\'eter S\'et\'any 1/A, Budapest, 1117, Hungary}

\author[0000-0001-5573-8190]{Zsolt Regály}
\affiliation{HUN-REN CSFK Konkoly Observatory, MTA Centre of Excellence, Konkoly Thege M. \'ut 15-17, Budapest, 1121, Hungary}



\begin{abstract}

Wolf--Rayet (WR) binaries frequently display high orbital eccentricities, and their eccentricity distributions are inconsistent with that of high-mass Galactic field binaries at high significance.
This raises the question of whether these eccentricities are primordial or generated during the WR phase.
We test whether a tidal enhancement of the mass-loss rate can pump the eccentricity of a WR binary on a sample of 14 WR systems.
We model each system as a variable-mass two-body problem and integrate the orbital evolution over the WR lifetime under three prescriptions: i) adiabatic isotropic mass loss, ii) equilibrium-tide circularization, and iii) a tidally enhanced, pericenter-focused mass loss in which the companion-facing reduction of the effective surface gravity raises the local Eddington parameter.
Adiabatic mass loss can not change eccentricity and affects only the orbital separation, which increases by up to a factor of $\sim20$.
Tidal circularization is effective only in the tightest orbits, which are compatible with the observed systems only for high mass-loss rates, a regime in which any primordial eccentricity is erased.
The tidally enhanced mass loss channel pumps the eccentricity by only $\leq10^{-4}$, three to four orders of magnitude below the observed values, and therefore cannot be the responsible mechanism.
We conclude that the high eccentricities of WR binaries are predominantly primordial rather than generated by mass loss during the WR phase.

\end{abstract}


\section{Introduction} \label{sec:intro}

Wolf--Rayet (WR) stars represent a short-lived, evolved phase of the most massive stars, distinguished by hydrogen-depleted surfaces and dense, fast, optically thick winds that produce their characteristic broad emission-line spectra \citep{crowther2007}. 
As they are believed to be the progenitors of Type Ib and Ic supernovae, their mass-loss histories set the pre-supernova structure and remnant masses.
As such, WR stars remain central to interpreting stripped-envelope core-collapse supernovae and the formation of massive black holes \citep{sander2020}. 

Binarity is common among WR stars, and wind--wind collisions yield phase-dependent radio-to-X-ray emission \citep{WhiteTuthill2024}.
In carbon-rich systems, the shocked interaction region of the colliding stellar winds can also enable episodic dust formation, spectacularly resolved in, e.g., WR~140 \citep{lau2022}.
Intriguingly, several WR binaries retain highly eccentric orbits, raising questions about tidal circularization and possible eccentricity pumping during prior interaction \citep{WhiteTuthill2024}. 
In such systems, periodic infrared brightening episodes may reflect orbital phase-dependent dust condensation in otherwise steady winds, or genuinely periodic modulation of the mass-loss rate itself, an ambiguity we aim to address below.

Infrared outbursts and expanding dust shells provide a direct tracer of the orbital phase at which dense, rapidly cooling material is produced \citep{Williams1990,lau2022,Lieb2025_WR140}. 
Recent interferometric orbit determinations for other dust-producing Wolf--Rayet binaries (e.g., WR~137) 
link dust production to specific orbital phases (namely, the outbursts happen when the stars are at the pericenter of their orbit). 
For WR 137, the dynamically measured component masses (see~Tab.\ref{tab:wr_binaries}) match binary-evolution tracks in which the WR star was stripped by its own radiatively driven wind rather than by binary mass transfer \citep{RichardsonEtAl2024_WR137}.
Furthermore, only a modest amount of wind material appears to have been accreted by the companion, consistent with its Oe classification.

Several physical mechanisms operate in eccentric Wolf-Rayet binary systems to produce an intrinsic enhancement of the mass-loss rate near the pericenter passage.
The first and most important mechanism is the tidal reduction of the effective surface gravity $g_\mathrm{eff}$ on the companion-facing hemisphere at pericenter.
This drives an increase in the mass-loss rate $\dot{M}$ concentrated on the inner hemisphere of the WR star, which rate is steeply sensitive to the effective escape velocity \citep{FriendAbbott1986} and scales steeply with the classical Eddington parameter $\Gamma_e = L/L_{\rm Edd}$ (the ratio of stellar luminosity $L$ and Eddington-luminosity $L_{\rm Edd}$) as $(1 - \Gamma_e)^{-n}$, where $n\approx3.5$ for optically thick winds \citep{SanderVinkHamann2020, Bestenlehner2020}. 

The second mechanism is the time-varying tidal potential excites stellar oscillation modes and induces shear flows in the outer envelope that peak near pericenter \citep{MorenoKoenigsbergerHarrington2011}.
\citet{KoenigsbergerMoreno2009, KoenigsbergerEstrellaTrujillo2024} applied this framework to the WR/luminous blue variable system HD~5980 and showed that tidal energy dissipation can trigger enhanced mass loss, which is expected to be non-uniform over the stellar surface.
\citet{Fuller2017} demonstrates that resonant coupling between orbital harmonics and stellar pulsation eigenmodes is strongly favored in high-eccentricity systems, where the pericenter passage spans a broad range of forcing frequencies and can lock onto a resonance that amplifies the tidal response, concentrating dissipation and any resulting mass loss near pericenter.
Note that tidally excited oscillations have actually been detected in the massive O-star binary $\iota$~Ori \citep{Pabloetal2017}.

The final mechanism, as the companion’s radiation field increases toward pericenter as $r^{-2}$, can modify the wind dynamics through radiative inhibition, which reduces wind acceleration toward the collision region \citep{StevensPollock1994}, and radiative braking, which slows the incoming wind before collision \citep{GayleyOwockiCranmer1997}.
These effects modify the ionization balance and the spectral-line opacity through which radiation transfers momentum to the wind, setting its mass-loss rate.

Of the three mechanisms discussed above, we focus on the first because it is the most direct and quantifiable. 
The oscillatory and radiative channels can also modulate the wind, but they are not easily included in an orbit-averaged secular framework and are left for future work.

In the orbital-dynamics sense, a quasi-steady WR wind can be treated as slow (the timescale of the mass loss is significantly larger than the orbital period of the binary), approximately isotropic mass loss relative to the orbital period, i.e., an adiabatic perturbation \citep{Hadjidemetriou1963, DosopoulouKalogera2016b, Veras2011}.  
In this limit, the semi-major axis evolves on the mass-loss timescale (widening as the system mass decreases), while the eccentricity is conserved to leading order.  
If instead mass and angular momentum are removed impulsively or with strong orbital-phase dependence, the osculating elements vary within an orbit, and the orbit-averaged eccentricity can grow (eccentricity pumping) \citep{DosopoulouKalogera2016b, SaladinoPols2019, regaly-etal-22}.  


Here we test whether continuous, adiabatic wind-driven mass loss, or a tidally enhanced, pericenter-focused, orbital-phase-dependent mass loss, can account for the large observed eccentricities.
We test the three-part argument that follows. 
First, by running the adiabatic evolution backward in time, we reconstruct the birth separations of observed WR binaries and quantify how much the orbit has expanded during the WR phase by solving a variable-mass two-body problem. 
Second, for systems whose inferred birth separations fall below the tidal circularization threshold -- where the equilibrium-tide damping timescale is shorter than the WR lifetime -- a primordial origin of the present-day eccentricity is ruled out, because any initial eccentricity would have been erased before the system reached its current state.
Third, we test whether an orbital phase-dependent, tidally enhanced mass loss that peaks at the pericenter can pump the eccentricity from a near-zero value up to the observed level within the available WR lifetime.
If it can, phase-dependent outflows are a highly important and necessary addition to the standard picture.

Following this introduction, Section 2 describes the binary sample, the adiabatic mass-loss model, the tidal evolution framework, and the pericenter-centered mass-loss prescription; Section 3 presents results for adiabatic mass loss, tidal evolution, and orbital phase-dependent mass loss; and Section 4 presents a discussion of our results. 
We present our conclusions in Section 5.

\section{Methods}

\subsection{Sample of WR binaries}

\begin{table*}[t]
\centering
\caption{Representative sample of binary systems with at least one WR component. 
The columns list component masses $M_{\mathrm{WR}}$ and $M_2$, physical radii $R_{\mathrm{WR}}$ and $R_2$, orbital semi-major axis $a$, orbital period $P$, and eccentricity $e$.
For most dust-producing systems, the dust morphology and/or orbital periodicity and eccentricity are known, while a complete dynamical solution (masses and $a$) is not yet available; however, these systems are also included for completeness.  
Five systems in the sample are, or may be, higher-order multiples (Apep, WR~11, WR\,48a, WR~138 and WR\,20a); the consequences of which are discussed in Section~\ref{sec:discussion}.
We treat hydrogen-rich WNha stars as Wolf–Rayet stars, following common usage; several primaries in our sample (WR 20a, WR 21a, WR 22, WR 25) are of this class.
}
\label{tab:wr_binaries}
\begin{tabular}{lllllllc}
\hline
System  & $M_{\rm WR}\,[M_\odot]$ & $M_{2}\,[M_\odot]$ & $R_{\rm WR}\,[R_\odot]$ & $R_{2}\,[R_\odot]$ & $a\,[{\rm au}]$ & $P\,[{\rm d}]$ & $e$ \\
\hline 
\multicolumn{8}{l}{Confirmed dust rings} \\
\hline
WR~140$^{1}$ &  $10.31\pm0.45$ & $29.27\pm1.14$ & \nodata & 35 & $13.55\pm0.21$ & $2895.0\pm0.3$ & $0.8993\pm0.0006$ \\
WR~137$^{2}$ &  $9.49\pm3.41$ & $17.34\pm1.91$ & $3.8\pm1$ & $7.7\pm1$ & $16.64\pm0.34$ & $4786.5\pm12.6$ & $0.3162\pm0.0023$ \\
Apep$^{3}$ &  \nodata & \nodata & \nodata & \nodata & \nodata & $70490\pm4020$ & $0.82\pm0.04$ \\
WR~125$^{4}$ &  \nodata & \nodata & \nodata & \nodata & \nodata & $10270$ & $0.29\pm0.12$ \\
WR~112$^{5}$ &  \nodata & \nodata & \nodata & \nodata & \nodata & $7310\pm40$ & $\simeq 0$ \\
WR~48a$^{6}$ &  \nodata & \nodata & \nodata & \nodata & \nodata & $\sim 11700$ & $0.6$ \\
\hline
\multicolumn{8}{l}{No dust rings} \\ \hline
WR~20a$^{7}$ &  $82.2\pm4.7$ & $81.4\pm4.7$ & $14.1\pm0.4$ & 
$19.3\pm0.5$ & $0.256$ & $3.686$ & $0$ \\
WR~11$^{8}$ &  $9\pm0.6$ & $28.5\pm1.1$ & $1.9$ & $16.2$ & $1.2$ & $78.524$ & $0.322\pm0.004$ \\
WR~139$^{9}$ &  $10.7$ & $26.4$ & $3$ & $8.5$ & $0.171$ & $4.212$ & $\simeq0$ \\
WR~21a$^{10}$ &  $94.4$ & $53.6$ & $23.3\pm1.6$ & $14.8\pm2.0$ & $1.070$ & $31.673$ & $0.695\pm0.007$ \\
WR~22$^{11}$  &  56.38 & 21.00 & $22.65$ & 11 & 1.535 & $80.34$ & 0.598 \\
WR~25$^{12}$  &  \nodata & \nodata & \nodata & \nodata & \nodata & $208$ & $0.50\pm0.02$ \\
WR~133$^{13}$ &  $9.3\pm1.6$ & $22.6\pm3.2$ & $3.4\pm0.5$ & $16.6\pm1.0$ & 1.45 & $112.8$ & $0.3646\pm0.0103$ \\
WR~138$^{14}$ &  13 & \nodata & 9.4 & \nodata & $5.81 \pm 0.13$ & $1521\pm35$ & 0.160 \\
\hline
\end{tabular}
\begin{flushleft}
\footnotesize
References.
$^{1}$~\citet{thomas2021wr140}; \citet{williamsperedur11}.
$^{2}$~\citet{RichardsonEtAl2024_WR137}.
$^{3}$~\citet{white2025apep}.
$^{4}$~\citet{richardson2024wr125}.
$^{5}$~\citet{Lau2020}.
$^{6}$~\citet{Williams2012}.
$^{7}$~\citet{rauwetal04}; \citet{schnurreetal09}.
$^{8}$~\citet{schmutzetal26}; \citet{north2007gamma2vel}.
$^{9}$~\citet{shaposhnikov2023v444cyg}.
$^{10}$~\citet{tramper2016wr21a}; \citet{gosset-nase16}; \citet{barbaetal22}.
$^{11}$~\citet{gosset2009wr22}; \citet{schweickhardtetal99}.
$^{12}$~\citet{gamen2006wr25}.
$^{13}$~\citet{richardsonetal21}.
$^{14}$~\citet{gvaramadzeetal09}.
\end{flushleft}
\end{table*}

\begin{figure}
    \centering
    \includegraphics[width=0.45\textwidth]{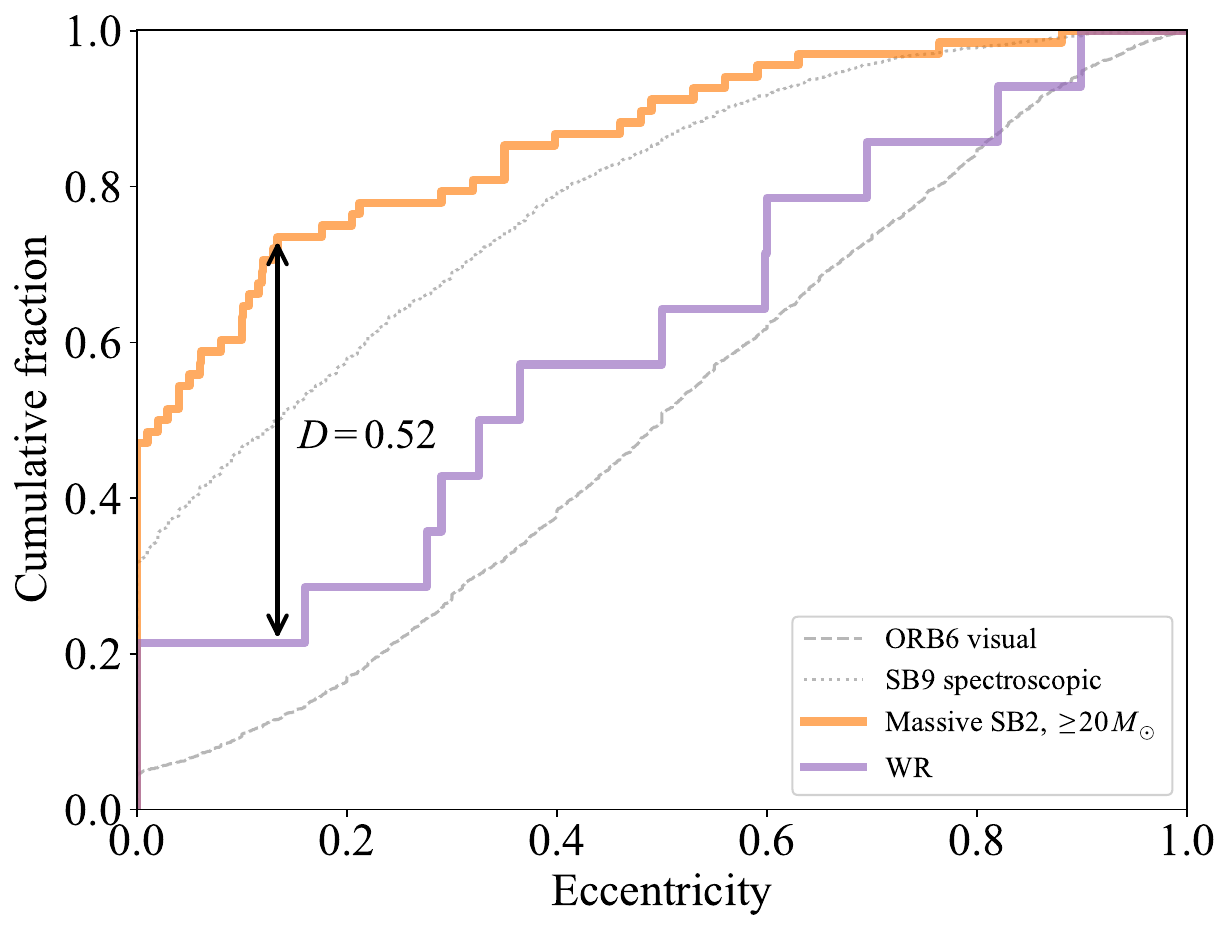}
    \caption{
    Empirical cumulative distribution functions of orbital eccentricity. The Wolf–Rayet binaries of Table 1 (purple) are compared against a mass-matched sample of SB9 double-lined systems with $(M_1+M_2)\sin^3 i > 20\,M_\odot$ (orange, N=68). The full SB9 spectroscopic (dotted, N=3160) and ORB6 visual (dashed, N=4045) catalogs are shown faintly for context. The double-headed arrow marks the Kolmogorov–Smirnov statistic $D=0.52$ (the maximum vertical separation) between the WR and mass-matched distributions. The WR systems are shifted toward markedly higher eccentricity than mass-matched field binaries (exact KS $p=2.0\times10^{-3}$; Scholz–Stephens AD $p=7\times10^{-4}$); the close agreement between the WR and ORB6 visual curves reflects a shared selection toward wide, dynamically preserved orbits rather than a physical similarity.
    }
    \label{fig:system-eccs}
\end{figure}

Table~\ref{tab:wr_binaries} summarizes a set of Wolf--Rayet binaries that are relevant to our study, divided into two groups: 
one in which dust structures have been detected, and another in which they have not.
The studied WR binaries are far more eccentric than comparable field binaries.
We tested this against reference samples from the Ninth Catalogue of Spectroscopic Binary Orbits (SB9, cross-matched with Gaia using the HEASARC \texttt{SBORBITCAT} table; \citealp{Pourbaix2004}) and the Sixth Catalog of Orbits of Visual Binary Stars (ORB6; \citealp{GaiaDR3Summary,Halbwachs2023, GaiaDR3NSS, ORB6, Hartkopf2001}).
Three sets were compared: 
1) the full SB9 spectroscopic sample ($N=3160$), 2) the full ORB6 visual sample ($N=4045$), and 3) a mass-matched subsample of SB9 double-lined systems with dynamical total mass $(M_1 + M_2)\sin^3 i > 20\,M_\odot$ ($N=68$).
Figure~\ref{fig:system-eccs} shows their empirical cumulative distribution functions alongside the WR sample of Table~\ref{tab:wr_binaries}.

We assess each comparison with the exact two-sample Kolmogorov--Smirnov test and the $k$-sample Anderson--Darling test of \citet{ScholzStephens1987}, the latter being more sensitive to differences in the distribution tails.
The mass-matched sample is the most important, since WR binaries descend from massive O+O progenitors.
Against it, the WR distribution is inconsistent with being drawn from the same parent population (compare the colored lines in Fig.~\ref{fig:system-eccs}; exact two-sample KS: $D=0.52$, $p = 2.0\times10^{-3}$; Anderson–Darling: $A^2 = 8.6$, $p = 7\times10^{-4}$).
The tail drives the contrast: only $\simeq 4\%$ of massive field binaries reach $e \geq 0.6$ (median $e = 0.03$, a signature of the strong tidal circularization of close spectroscopic pairs), against $\simeq 29\%$ of the WR systems (median $e = 0.35$).
Dropping the two circularized WR systems ($e \simeq 0$) sharpens the discrepancy further (KS: $D=0.66$, $p = 8.0\times10^{-5}$; AD: $A^2 = 13.7$, $p = 3\times10^{-5}$).

The ORB6 distribution is itself skewed toward high eccentricity (median $e=0.50$) and closely resembles the WR sample.
We read this as a selection effect, not a physical similarity: 
visual resolution preferentially recovers wide, long-period orbits, in which any primordial eccentricity survives.
We note, however, that the mass-matched sample is defined by $(M_1 + M_2)\sin^3 i$, a lower bound on the true mass, and, as a spectroscopic sample, it carries the same short-period circularization bias as SB9 as a whole.
Both effects lower the field eccentricities relative to a truly mass- and separation-matched comparison, so both are conservative for our argument.
As such, high eccentricities are rare among binaries as massive as our studied set of WR systems.

\subsection{Adiabatic mass loss}
\label{sec:methods:adiabatic}

First, we investigate the orbital evolution of eccentric Wolf--Rayet binaries under adiabatic mass loss.
In this limit, the mass decreases slowly compared to the orbital period, leading primarily to secular widening of the orbit while conserving eccentricity \citep{Hadjidemetriou1963,Veras2011}.

We thus model the orbital response using the standard adiabatic and isotropic mass-loss approximation (e.g. \citealt{Veras2011}), which applies when the characteristic mass-loss timescale
is long compared to the orbital period.
Here the system evolves through a sequence of osculating Keplerian orbits, and short-timescale phase dependencies average out. 
The perturbation can be recast as an additional acceleration proportional to the velocity and the fractional mass-loss rate, preserving rotational symmetry and hence angular momentum, while allowing the orbital energy to vary \citep{Hadjidemetriou1963}. 
In the adiabatic limit, the eccentricity remains (nearly) constant, and the secular evolution reduces to the invariant $a\,\mu \simeq \mathrm{const}$, where $\mu(t)\equiv G\,[M_1(t)+M_2(t)]$ is the time-dependent gravitational parameter.

We sample $\dot{M}_0 \in \{10^{-6}, 10^{-5}, 10^{-4}\}\, M_\odot\,\mathrm{yr^{-1}}$ in all of the simulations in this paper, which brackets the empirically derived range for Galactic and Magellanic Cloud classical WR stars \citep[$10^{-5.5} \lesssim \dot{M} \lesssim 10^{-4.5}\, M_\odot\,\mathrm{yr^{-1}}$;
see][]{Hamannetal2019, sander2020}.
For the day--year orbital periods characteristic of many WR binaries and colliding-wind systems, 
the orbit-averaged, adiabatic treatment captures the leading secular effect of smoothly decreasing $\mu(t)$ and hence expanding $a(t)$. 
We note that the exact value of $\dot{M}$ depends on metallicity and the Eddington factor $\Gamma=L/L_{\rm Edd}$. 


The mass loss prescription in the adiabatic regime thus yields a change in semi-major axis equivalent to
\begin{equation}
\left(\frac{da}{dt}\right)_{\rm ad}
=
-\frac{a}{\mu}\,\frac{d\mu}{dt}
=
-~a~\frac{\dot M_1+\dot M_2}{M_1+M_2},
\label{eq:dadt_adiabatic}
\end{equation}
as in our calculations, the companion star does not undergo mass loss; hence $\dot M_2 = 0$ is assumed.

\subsection{Equilibrium tides with a constant time lag (CTL) under pseudo-synchronization}
\label{sec:methods:tidal}

We adopt the equilibrium-tide weak friction formalism with a constant time lag (CTL), as developed by \citet{Hut1981}.
We restrict to aligned spins (zero obliquity) and treat both stellar components symmetrically by superposition of the tides raised on each star.
To eliminate the (generally unknown) stellar spin frequencies $\Omega_i$ from the orbital evolution, we close the system by assuming that both stars have reached the CTL pseudo-synchronous state, i.e., the orbit-averaged tidal torque on each star vanishes ($d\Omega_i/dt=0$). 
This is justified as for stellar-mass companions spin–orbit pseudo-synchronization proceeds orders of magnitude faster than tidal circularization \citep{Zahn1977, Hut1981}, so the spin reaches its equilibrium value long before the orbit evolves appreciably.

Consider a binary with component masses $M_i$, radii $R_i$, and spin angular frequencies $\Omega_i$ ($i\in\{1,2\}$; $j\equiv 3-i$),
orbital semi-major axis $a$, eccentricity $e$, and mean motion
$n = \left[G~(M_1+M_2)~/~a^3\right]^{1/2}$.
For each star, we adopt a degree-2 Love number $k_{2,i}$ and a constant tidal time lag $\Delta t_i$.
In the notation of \citet{Hut1981}, it is convenient to define
\begin{equation}
T_i \equiv \frac{R_i^{3}}{G M_i \Delta t_i},
\qquad
q_i \equiv \frac{M_j}{M_i}.
\end{equation}

In the CTL equilibrium-tide framework, tidal dissipation is parameterized by a single lag time $\Delta t$ between the line of centers and the tidal bulge response (e.g. \citealt{Hut1981,EggletonKiselevaHut1998}). 
However, stellar theory and observations both indicate that the effective dissipation is, in general, frequency dependent and may vary by orders of magnitude across stellar types and forcing periods (see the review by \citealp{Ogilvie2014}). 
A practical way to set $\Delta t$ is to calibrate it to the empirically inferred modified tidal quality factor $Q$ \citep{PatelPenevSchussler2023}. 
In this spirit, and to keep the parameter space tractable, we adopt a fiducial $\Delta t$ corresponding to an intermediate regime, $\Delta t=0.1~{\rm s}$, noting that this value has been used successfully in population-level synchronization modeling of Kepler and TESS eclipsing binaries \citep[e.g.][]{Crawford2025,Fleming2019,Lurie2017}.

Regarding the degree-2 Love number $k_2$, 
we adopt $k_2$ from stellar-structure grids computed via integration of the Radau equation along evolutionary tracks.
Modern tabulations show that on the main sequence, typical values are $k_2\simeq 10^{-2}$ for $\simeq 1\,M_\odot$ stars and $k_2\simeq 10^{-3}$ for several-$M_\odot$ early-type stars, with $k_2$ generally decreasing as stars become more centrally condensed and evolve off the main sequence \citep{Claret2023}.
Measurements in well-characterized double-lined eclipsing binaries provide empirical constraints that broadly agree with predictions \citep{ClaretGimenezEtAl2021}. 
As WR stars are among the most massive early-type stars, we adopt $k_2=10^{-3}$.

For tides raised on star $i$ and averaged over the orbit, the CTL equations for $a$ and $e$ can be written as \citep[][his Eqs.~9--10]{Hut1981}
\begin{align}
\left(\frac{da}{dt}\right)_i
&=\notag
-6\,\frac{k_{2,i}}{T_i}\,q_i(1+q_i)\left(\frac{R_i}{a}\right)^{8}
\frac{a}{(1-e^2)^{15/2}}\times\\
&\times\left[
f_1(e^2) - (1-e^2)^{3/2} f_2(e^2)\frac{\Omega_i}{n}
\right],
\label{eq:ctl_dadt_i}
\end{align}

\begin{align}
\left(\frac{de}{dt}\right)_i
&=\notag
-27\,\frac{k_{2,i}}{T_i}\,q_i(1+q_i)\left(\frac{R_i}{a}\right)^{8}
\frac{e}{(1-e^2)^{13/2}}\times \\
&\times\left[
f_3(e^2) - \frac{11}{18}(1-e^2)^{3/2} f_4(e^2)\frac{\Omega_i}{n}
\right].
\label{eq:ctl_dedt_i}
\end{align}
The total orbital evolution is the sum of both contributions,
\begin{equation}
\frac{da}{dt}=\sum_{i=1}^{2}\left(\frac{da}{dt}\right)_i,
\quad
\frac{de}{dt}=\sum_{i=1}^{2}\left(\frac{de}{dt}\right)_i .
\end{equation}
The eccentricity functions are 
\begin{align}
f_1(e^2) &= 1 + \frac{31}{2}e^2 + \frac{255}{8}e^4 + \frac{185}{16}e^6 + \frac{25}{64}e^8, \\
f_2(e^2) &= 1 + \frac{15}{2}e^2 + \frac{45}{8}e^4 + \frac{5}{16}e^6, \\
f_3(e^2) &= 1 + \frac{15}{4}e^2 + \frac{15}{8}e^4 + \frac{5}{64}e^6, \\
f_4(e^2) &= 1 + \frac{3}{2}e^2 + \frac{1}{8}e^4, \\
f_5(e^2) &= 1 + 3e^2 + \frac{3}{8}e^4.
\end{align}

In the CTL model, the pseudo-synchronous spin rate for aligned spins is \citep[][his Eq.~42]{Hut1981}
\begin{equation}
\Omega_{\rm ps}
=
n~\frac{f_2(e^2)}{f_5(e^2)\,(1-e^2)^{3/2}}.
\label{eq:ctl_omega_ps}
\end{equation}
Assuming both stars have relaxed to pseudo-synchronization, $\Omega_i=\Omega_{\rm ps}(e)$, which eliminates the spins from
Eqs.~(\ref{eq:ctl_dadt_i})--(\ref{eq:ctl_dedt_i}). Using Eq.~(\ref{eq:ctl_omega_ps}),
the closed orbital evolution becomes
\begin{align}
\frac{da}{dt}
&=\notag
-6\,\frac{a}{(1-e^2)^{15/2}}
\left[
f_1(e^2)-\frac{f_2(e^2)^2}{f_5(e^2)}
\right]\times\\
&\times\sum_{i=1}^{2}
\frac{k_{2,i}}{T_i}\,q_i(1+q_i)\left(\frac{R_i}{a}\right)^{8},
\label{eq:ctl_dadt_ps}
\\[4pt]
\frac{de}{dt}
&=\notag
-27\,\frac{e}{(1-e^2)^{13/2}}
\left[
f_3(e^2)-\frac{11}{18}\frac{f_4(e^2)f_2(e^2)}{f_5(e^2)}
\right]\times\\
&\times\sum_{i=1}^{2}
\frac{k_{2,i}}{T_i}\,q_i(1+q_i)\left(\frac{R_i}{a}\right)^{8}.
\label{eq:ctl_dedt_ps}
\end{align}

We define the semi-major-axis evolution and eccentricity damping timescales as
$\tau_a \equiv \left|{a}/{\dot a}\right|$ and
$\tau_e \equiv \left|{e}/{\dot e}\right|$.
Using Eqs.~(\ref{eq:ctl_dadt_ps})--(\ref{eq:ctl_dedt_ps}), these can be expressed as 
\begin{align}
\tau_a^{-1}
&=\notag
6\,\frac{1}{(1-e^2)^{15/2}}
\left|
f_1(e^2)-\frac{f_2(e^2)^2}{f_5(e^2)}
\right|\times\\
&\times\sum_{i=1}^{2}
\frac{k_{2,i}}{T_i}\,q_i(1+q_i)\left(\frac{R_i}{a}\right)^{8},
\label{eq:ctl_tau_a_ps}
\\[4pt]
\tau_e^{-1}
&=\notag
27\,\frac{1}{(1-e^2)^{13/2}}
\left|
f_3(e^2)-\frac{11}{18}\frac{f_4(e^2)f_2(e^2)}{f_5(e^2)}
\right|\times\\
&\times\sum_{i=1}^{2}
\frac{k_{2,i}}{T_i}\,q_i(1+q_i)\left(\frac{R_i}{a}\right)^{8}.
\label{eq:ctl_tau_e_ps}
\end{align}

\subsection{Orbit-averaged secular evolution under tidally enhanced mass loss}\label{sec:methods:orbit_averaging}

Here we describe the framework we employ to evolve the
orbital elements of the WR binary systems under the orbital phase-dependent mass loss assumption.
This prescription, as opposed to the adiabatic one, can modify the osculating elements within an orbit and potentially excite eccentricity \citep{DosopoulouKalogera2016a, DosopoulouKalogera2016b, Sepinskyetal2007, Sepinskyetal2009, regaly-etal-22}.
We treat the binary as a Keplerian two-body system perturbed by mass loss from the WR component.
Throughout this work, we adopt the Jeans-mode approximation, in which each mass element leaves the system carrying the local specific orbital energy and angular momentum of the donor star at the point of ejection \citep{HuangEtAl1956, HadjiDemetriou1969}.
This is appropriate for fast stellar winds with
$v_\mathrm{wind} \gg v_\mathrm{orbital}$, a condition comfortably
satisfied by WR winds with $v_\infty \simeq 1000$--$3000\,\mathrm{km\,s^{-1}}$
and orbital velocities of at most a few hundred $\mathrm{km\,s^{-1}}$
even at pericenter.

Under the Jeans-mode assumption, the instantaneous perturbation
to the semi-major axis $a$ and eccentricity $e$ produced by an
infinitesimal mass loss $dM_{\rm tot}$
at true anomaly $\nu$ is
\citep[e.g.,][]{HadjiDemetriou1969, Sepinskyetal2007, DosopoulouKalogera2016a, DosopoulouKalogera2016b}
\begin{equation}
  \frac{da}{a}    = -\left(\frac{2a}{r(\nu)} - 1\right)
                     \frac{dM_{\rm tot}}{M_{\rm tot}}\,, \label{eq:da_inst}\
\end{equation}
\begin{equation}
  de              = -\left(\cos\nu + e\right)
                     \frac{dM_{\rm tot}}{M_{\rm tot}}\,, \label{eq:de_inst}
\end{equation}
where $M_{\rm tot} = M_{\rm WR} + M_2$ is the total system mass and
$r(\nu) = a(1-e^2)/(1 + e\cos\nu)$ is the instantaneous separation.

The fractional mass lost per orbital period is small for all systems considered in this work.
This means one can use the standard secular framework where 
orbital elements are time-averaged over one period of
the unperturbed Keplerian motion
\citep[see][for a comprehensive derivation]{DosopoulouKalogera2016a, DosopoulouKalogera2016b}.
The time-average of an arbitrary phase-dependent function $f(\nu)$
over one orbital period $P_\mathrm{orb}$ is
\begin{equation}
\begin{split}
  \left\langle f \right\rangle_t
  \;&=\;
  \frac{1}{P_\mathrm{orb}} \int_0^{P_\mathrm{orb}} f(\nu(t))\, dt=\\
  \;&=\;
  \frac{(1 - e^2)^{3/2}}{2\pi}
  \int_0^{2\pi} \frac{f(\nu)}{\bigl(1 + e\cos\nu\bigr)^2}\, d\nu\,.
  \label{eq:time_average}
\end{split}
\end{equation}
%

Applying Eq.~(\ref{eq:time_average}) to
Eqs.~(\ref{eq:da_inst})--(\ref{eq:de_inst}), with the
mass-loss rate 
$\dot{M}$
where $\dot{M}_0$ is the intrinsic mass-loss rate
and $\xi(\nu)$ is a dimensionless phase-dependent enhancement
factor with $\langle\xi(\nu)\rangle = 1$ for an isotropic
constant wind, the orbit-averaged secular rates become
\begin{align}
  \left\langle \dot{M}_{\rm WR} \right\rangle &=
    -\dot{M}_0\, \big\langle \xi(\nu) \big\rangle_t\,,
    \label{eq:dMdt_avg}\\[0.4em]
  \left\langle \frac{\dot{a}}{a} \right\rangle &=
    \frac{\dot{M}_0}{M_{\rm tot}}\,
    \left\langle\!
      \left(\frac{2a}{r} - 1\right) \xi(\nu)
    \!\right\rangle_t\,,
    \label{eq:dadt_avg}\\[0.4em]
  \left\langle \dot{e} \right\rangle &=
    \frac{\dot{M}_0}{M_{\rm tot}}\,
    \big\langle (\cos\nu + e)\, \xi(\nu) \big\rangle_t\,.
    \label{eq:dedt_avg}
\end{align}
Note that it is clear from Eq.~(\ref{eq:dedt_avg}) that for an isotropic, constant wind $\langle \cos\nu + e \rangle_t = 0$, which is the \citet{Jeans1924} adiabatic result.
Any phase dependence of $\xi(\nu)$ that breaks this symmetry
produces a nonzero secular
$\langle \dot{e}\rangle$.



We now apply this framework to the most efficient channel for intrinsic mass-loss enhancement: the tidally driven reduction of the effective surface gravity.
A star of mass $M_{\rm WR}$, radius $R_{\rm WR}$ in the field of a companion $M_2$ at separation $r$ develops a tidal bulge. 
The equilibrium shape is an oblate/prolate ellipsoid whose oblateness is quantified by the dimensionless tidal deformation parameter $\varepsilon$ \citep{Love1911, Kopal1959}:
\begin{equation}
    \varepsilon_{\rm T}=\frac{\Delta R}{R_{\rm WR}}=k_2~\frac{M_2}{M_{\rm WR}}~\left( \frac{R_{\rm WR}}{r}\right)^3.
\end{equation}

This tidal distortion is going to cause the effective gravitational acceleration on the surface to be lower, which in turn drives an increase in $\dot M$ that scales with the Eddington parameter as \citep{Bestenlehner2020}
\begin{equation}\label{eq:mdot_gamma}
  \dot{M} \;\propto\; \frac{1}{\bigl(1 - \Gamma_e\bigr)^{n}}\,,
\end{equation}
where $n \approx 3.5$ for optically thick winds
(see later).
The classical Eddington parameter is defined as
\begin{equation}\label{eq:gamma_e}
  \Gamma_{e,0} \;=\; \frac{L}{L_{\mathrm{Edd}}}\, = \frac{\kappa_e \, L_{\rm WR}}{4\pi \, c \, G \, M_{\rm WR}},
\end{equation}
where $\kappa_e$ is the electron-scattering opacity and $L_{\rm WR}$ and $M_{\rm WR}$ is the stellar luminosity and mass, respectively.

At orbital separation $r$, the tidal perturbation to the surface gravity on the companion-facing point is, to leading order,
\begin{equation}\label{eq:geff}
  g_{\mathrm{eff}} \;=\; g_0 \left(1 \;-\; \varepsilon_{\rm T}\right)\,,
  \qquad
\end{equation}
where $g_0 = G M_{\rm WR} / R_{\rm WR}^2$ is the unperturbed surface gravity.
In an eccentric orbit $r = a(1 - e\cos E)$, where $E$ is the eccentric anomaly, so $\varepsilon_{\rm T}$ peaks at pericenter where $r = a(1-e)$.
The tidal reduction of $g_{\mathrm{eff}}$
maps onto an increased local Eddington parameter.
Replacing $g_0$ by $g_{\mathrm{eff}}$ in the definition of $\Gamma_{e,0}$ yields
\begin{equation}\label{eq:gamma_tidal}
  \Gamma_e
  \;=\; \frac{\Gamma_{e,0}}{1 - \varepsilon_{\rm T}}.
\end{equation}
Inserting Eq.~(\ref{eq:gamma_tidal}) into
Eq.~(\ref{eq:mdot_gamma}), the tidally enhanced mass-loss rate
becomes
\begin{align}
  \dot{M}_{\rm WR} = \dot M_0 ~\xi(\nu)
  \;&=\; \dot{M}_{\rm 0} \;
  \left[
    \frac{(1 - \Gamma_{e,0})(1 - \varepsilon_{\rm T})}
         {1 - \varepsilon_{\rm T} - \Gamma_{e,0}}
  \right]^{n}\,,
\label{eq:mdot_peri}
\end{align}
We note that centrifugal effects from stellar rotation would further reduce the effective gravity and enhance $\dot{M}$, making our estimates conservative. 

Substituting Eq.~(\ref{eq:mdot_peri})
into the orbit-averaged secular rates
of Eqs.~(\ref{eq:dMdt_avg})--(\ref{eq:dedt_avg}) yields a
closed-form prescription for
$\langle \dot{M}_{\rm WR}\rangle$,
$\langle \dot{a}\rangle$, and
$\langle \dot{e}\rangle$
that depends only on the orbital elements, the component
masses, the WR radius, and the wind
parameters $\Gamma_{e,0}, n, k_2,$ and $\dot{M}_0$.

The orbit-averaging integrals of Eqs.~(\ref{eq:dMdt_avg})-(\ref{eq:dedt_avg})
are evaluated numerically using adaptive Gauss-Kronrod quadrature
(\texttt{scipy.integrate.quad}, \citealp{2020SciPy-NMeth}) with a relative tolerance of $10^{-12}$.
The coupled system of secular ODEs,
is integrated using an explicit eighth-order Runge––Kutta method, \texttt{DOP853} \citep{HairerNorsettWanner1993}, as implemented in
\texttt{scipy.integrate.solve\_ivp}, with relative and absolute
tolerances of $10^{-10}$ and $10^{-13}$ respectively (note that this integrator has been used successfully for studying the dynamical consequences of mass loss in supernova systems in e.g., \citealp{regaly-etal-22,frohlich-etal-23,frohlich-regaly-25}).

For each system in our sample, we take the present-day observed orbital and stellar parameters as the initial condition, and integrate the secular ODE system backward in time, which means that the WR star regains the mass it has lost, the orbit contracts, and the eccentricity decreases.
We integrate each system backward to 
$-\tau_\mathrm{WR} \in -\{1, 5, 10, 50\} \times 10^5$~yr in order to
sample the range of plausible WR-phase durations and to place upper
bounds on the cumulative effect over timescales exceeding the
canonical WR lifetime.
Note that we halt
integration if $M_{\rm WR}$ exceeds $300\,M_\odot$, the empirical upper limit on stellar masses \citep{Crowther2010}.
For the longest backward integrations and the highest mass-loss rates, the upper-mass guard is sometimes triggered well before the nominal endpoint, indicating that the imposed history is internally inconsistent; we retain these runs as upper limits on the change in eccentricity.

The two key parameters in Eq.~(\ref{eq:mdot_peri}) are the thick-wind exponent $n$ and the unperturbed Eddington parameter $\Gamma_{e,0}$.
For the exponent, \citet{Bestenlehner2020} found empirically that $\dot{M} \propto (1 - \Gamma_e)^{-3.56 \pm 0.28}$ for the optically thick winds of WNh stars in R136, i.e. $n \approx 3.5$.
This value is consistent with the theoretical prediction of \citet{Vinketal2011}, whose models yield an exponent of 3.99, and with the empirical $\Gamma$-dependent mass-loss relation of \citet{Grafeneretal2011}, which matches this slope precisely.
We adopt $n=3.5$ as our fiducial value, noting that the \citet{Vinketal2011} exponent of $\approx 4$ would only strengthen the nonlinear enhancement.
For hydrogen-depleted classical WR stars (WN, WC), \citet{SanderVinkHamann2020} showed that the mass-loss rate depends on the Eddington parameter through a qualitatively similar but distinct scaling; adopting $n=3.5$ for these objects should therefore be regarded as approximate.

The unperturbed Eddington parameter $\Gamma_{e,0}$ is set by the WR star's luminosity-to-mass ratio and can be determined for each system individually through spectral modeling.
\citet{GrafenerHamann2008} identified WNL stars as very massive stars close to the Eddington limit, with their self-consistent atmosphere models for WR~22 yielding $\Gamma_e = 0.55$.
\citet{Grafeneretal2011} found that the empirical $\Gamma_e$ values for the WNh stars in the Arches cluster and NGC~3603 span the range $\Gamma_{e,0} \approx 0.5-0.7$, while the R136 sample of \citet{Bestenlehner2020} extends up to $\Gamma_e \approx 0.8-0.9$ for the most luminous WNh objects, with the transition from optically thin to thick winds occurring at $\Gamma_{e,\mathrm{trans}} \approx 0.47$.
For the calculations in this work, we explore $\Gamma_{e,0}$ in the range $0.5-0.9$, which brackets the observed values for Galactic and LMC WR stars.

The companion mass $M_2$ is held fixed throughout the integration on the grounds that the O-type companion experiences mass-loss rates one to two orders of magnitude lower than the WR primary \citep{VinkdeKoterLamers2001}.
We hold the stellar radius fixed throughout the integration, neglecting the modification of $R_{\rm WR}$ by the tidal deformation itself. 
Since $\varepsilon_{\rm T} \propto (R_{\rm WR}/r)^3$ and $R_{\rm WR}/r \ll 1$ for all systems even at pericenter, the back-reaction of the deformation on the radius is a higher-order correction and does not affect the mass-loss enhancement at the level of our approximation. The same holds for the tidal-circularization calculation of Sect~\ref{sec:methods:tidal}.

The integrable subsample comprises those WR binaries for which all five required parameters ($M_\mathrm{WR}$, $M_2$, $R_\mathrm{WR}$, $a$, $e$) have published values.
However, if a system is only missing the $R_\mathrm{WR}$ value, we assume it to be $1~R_\odot$ as the radial extent will not result in qualitative differences on the range of radii seen in Table~\ref{tab:wr_binaries}.

\section{Results}


\subsection{Adiabatic mass loss with backward-integration}
\label{sec:res:adiabatic}

Here we assume that WR binary systems undergo adiabatic, isotropic mass-loss for $\tau_{\mathrm{WR}}$ time.
Using the methods presented in Section~\ref{sec:methods:adiabatic}, we calculate the semi-major axis of systems at their birth by solving Eq.~\ref{eq:dadt_adiabatic}.
We sample the mass loss rate as $\dot M \in [10^{-6}-10^{-4}]~M_\odot~\rm yr^{-1}$.
We emphasize that adiabatic mass-loss can introduce a significant change in semi-major axis, but leaves the eccentricity of the system untouched.
As the integration is done backwards in time, the WR mass grows and the semi-major axis contracts in our calculations.

\begin{figure}
    \centering
    \includegraphics[width=0.99\columnwidth]{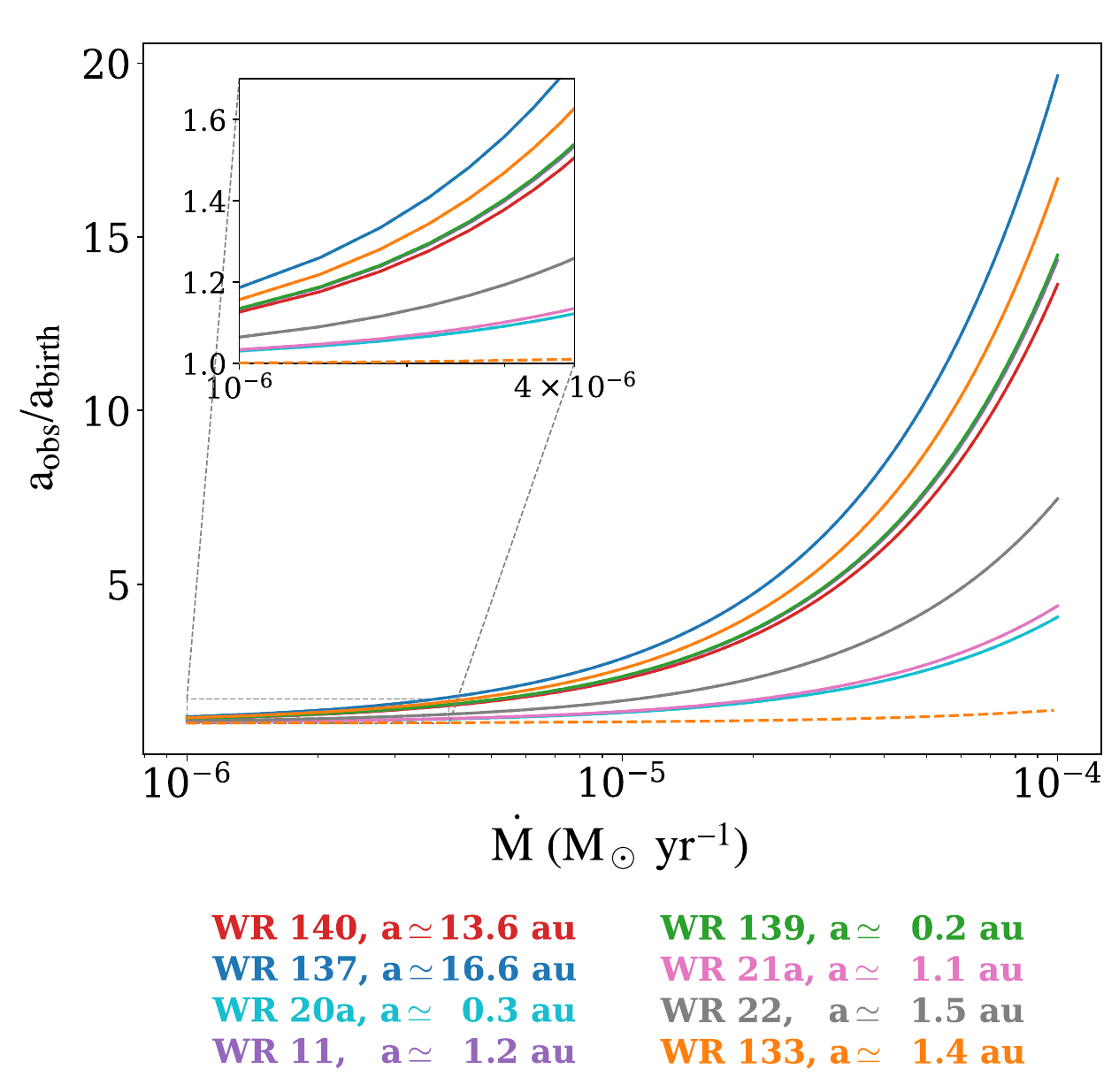}
    \caption{
    Ratio of initial and currently observed semi-major axes in the WR systems with known component masses and semi-major axes as a function of WR mass loss rate in the adiabatic model.
    Solid and dashed lines show extrema for WR lifetime, namely $5\times10^6$ and $5\times10^5$ years, respectively.
    The inset is a zoom on the low mass loss rate region.
    Systems are represented by colors; see legend.
    }
    \label{fig:adiabatic-v2}
\end{figure}

Figure~\ref{fig:adiabatic-v2} shows the growth in semi-major axis as a function of mass loss rate for the systems listed in Table~\ref{tab:wr_binaries}.
Note that systems that have a missing component mass or semi-major axis are not included.
Solid and dashed lines represent $\tau_{\rm WR}=5\times10^6$ and $5\times10^5$~years, respectively.
As one can see, a longer WR lifetime (larger $\tau_{\mathrm{WR}}$) and a higher mass-loss rate result in a larger change in separation, with the semi-major axis growing more and more over time.

Assuming $\tau_{\rm WR}=5\times10^5~\rm yr$, the change in semi-major axis is $\simeq1$ for all systems (see dashed lines), meaning that there is no visible growth in the semi-major axis. 
As WR lifetime increases, the fates of the examined systems diverge.
The change in $a$ is going to be inversely proportional to the total mass of the system ($M_{\rm WR}+M_{2}$) and is smallest for the most massive systems (compare, e.g., WR~137, where the total mass is $26.8~M_\odot$ and WR~20a, where the total mass is $163.6~M_\odot$).
As such, the change in semi-major axis may be as great as $a_{\rm obs}/a_{\rm birth}\simeq20$ for the least massive binary systems.
Looking at the physical separation of the systems at birth, a high mass loss rate of $\dot{M}=10^{-4}~M_\odot~\mathrm{yr^{-1}}$ may even imply $a_{\mathrm{birth}}$ close to the range of $10^{-3}-10^{-2}~\rm au$. 

\subsection{Tidal circularization with backward integration}
\label{sec:res:tidal-circ}

As the system evolves backward in time, it gains mass adiabatically, so its semi-major axis shrinks and tidal forces strengthen. Circularization may therefore dominate the early orbital evolution: where the circularization timescale falls well below the WR lifetime ($\tau_{\mathrm{tidal}}\ll\tau_{\mathrm{WR}}$), any primordial eccentricity would have been erased, and the observed eccentricity must have been pumped by another mechanism.

\begin{figure}
    \centering
    \includegraphics[width=0.99\columnwidth]{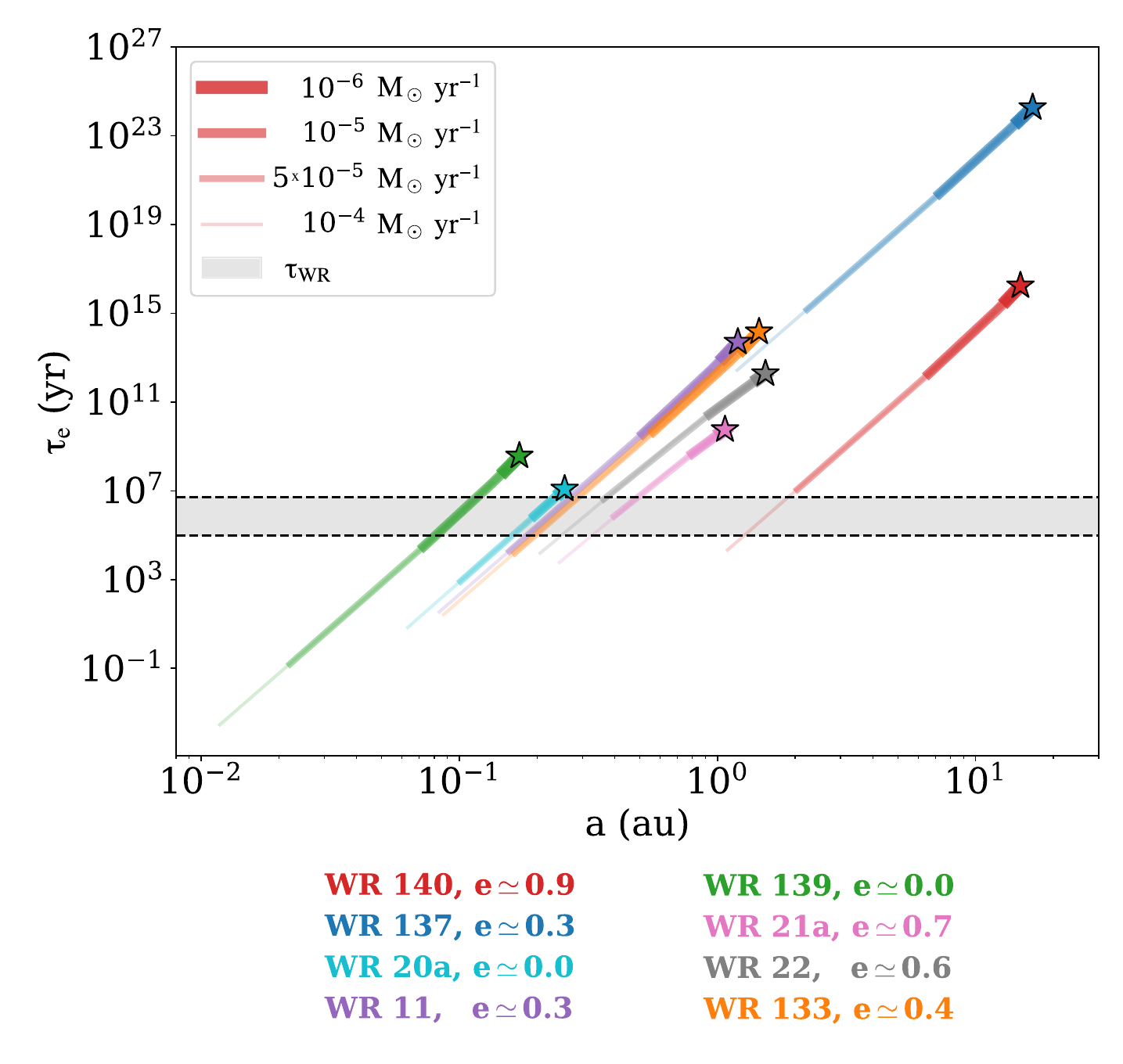}
    \caption{
    Tidal eccentricity timescale in representative WR systems.
    Stars mark the $\tau_{\rm e}$ corresponding to the current state.
    Each line traces $\tau_e$ evaluated at the reconstructed birth separation as the assumed WR lifetime is varied over $\tau_{\rm WR} \in \{10^5,\,5\times10^5,\,10^6,\,5\times10^6\}\,{\rm yr}$ at a fixed mass-loss rate, the far end of each line corresponding to the longest lifetime.
    Mass loss rates are denoted by different line width, among which different $\tau_{\rm WR}$ values are used. 
    Shorter lines belong to smaller mass loss rates, as in that case the change in the orbital parameters, and hence in $\tau_{\rm e}$, will be smaller.
    The currently observable eccentricity of the systems is noted in the bottom legend to the side of the figure.
    The gray band represents the average WR lifetime, namely $\tau_{\rm WR}\in[10^ {5}, 5\times10^6]$~years. 
    Note that systems whose component mass or semi-major axis is unknown are excluded, and those where the stellar radius is unknown are assumed $R=1~R_\odot$, as according to our calculations, the radius does not alter the results qualitatively, as the tidal effect scales with $(R_{\rm WR}/r)^3\ll1$.
    }
    \label{fig:tidal-timescales}
\end{figure}

Figure~\ref{fig:tidal-timescales} shows the tidal timescale $\tau_{\rm e}$ (Eq.~\ref{eq:ctl_tau_e_ps}) of the systems listed in Table~\ref{tab:wr_binaries} as a function of the currently observed semi-major axis of the systems, with mass loss rates denoted by different line widths.
A grey band shows the typical WR lifetime, $\tau_{\rm WR}\in[10^5-5\times10^6]$~years.
The birth separations reconstructed using the adiabatic model themselves depend on the assumed WR lifetime $\tau_{\rm WR}$ through the backward integration.
Therefore, each line in Fig.~\ref{fig:tidal-timescales} spans the full range $\tau_{\rm WR} \in \{10^5, 5\times10^5, 10^6, 5\times10^6\}\,{\rm yr}$: 
the star marks the present-day state (observed separation), while progressively larger $\tau_{\rm WR}$ moves the system back to smaller birth separations and correspondingly shorter $\tau_e$, tracing out the line.

As one can clearly see in Fig.~\ref{fig:tidal-timescales}, higher mass loss rates and longer WR timescales (see various line widths) lead to a more significant decrease in semi-major axis, which causes the tidal timescale of the system to also shorten.
For some systems, at the end of the integration time, this can mean that the tidal timescale can become much less than the WR timescale (these systems fall below the grey band on the figure).
This is true especially for WR~139, WR~20a, WR~11, and WR~133, but can also be observed in WR~22, WR~21a, and WR~140 for the most extreme mass loss rates and WR timescales.
As such, the eccentricity of these systems would have become zero so quickly that if they do possess eccentricity currently, it can not be primordial.

Notice systems WR~20a and WR~139, both of which are close to the $\tau_{\mathrm{tidal}}=\tau_{\mathrm{WR}}$ limit already at the start of the integration.
This is well reflected in their eccentricities, as currently $e\simeq0$ for both systems.
Other than these, systems require a mass loss rate of at least $\dot M\gtrsim5\times10^{-5}~M_\odot~\rm yr^{-1}$ to reach the $\tau_{\rm tidal}\lesssim\tau_{\rm WR}$ regime, that is, to circularize within the WR timescale.
This implies that a non-primordial origin for the eccentricity requires mass-loss rates at the upper end of the observed range.

Regarding WR~137, even the mass loss rate of $10^{-4}~M_\odot~\rm yr^{-1}$ can not bring the system into the $\tau_{\rm tidal}\ll\tau_{\rm WR}$ regime.
This means that the moderate eccentricity of this system ($e\simeq0.3$) is very likely primordial.
As for the rest of the investigated systems, eccentricity can only be primordial if the mass loss rate is $\lesssim 5\times10^{-5}~M_\odot~\rm yr^{-1}$ in the case of WR~11, WR~133, and $\lesssim 10^{-4}~M_\odot~\rm yr^{-1}$ for WR~140, WR~21a and WR~22.
With greater mass loss rates, one may assume that there is an eccentricity pumping mechanism at play.

\subsection{Tidally enhanced, orbital phase-dependent mass loss}
\label{sec:res:tidal-mass-loss}

\begin{figure}
    \centering
    \includegraphics[width=0.999\linewidth]{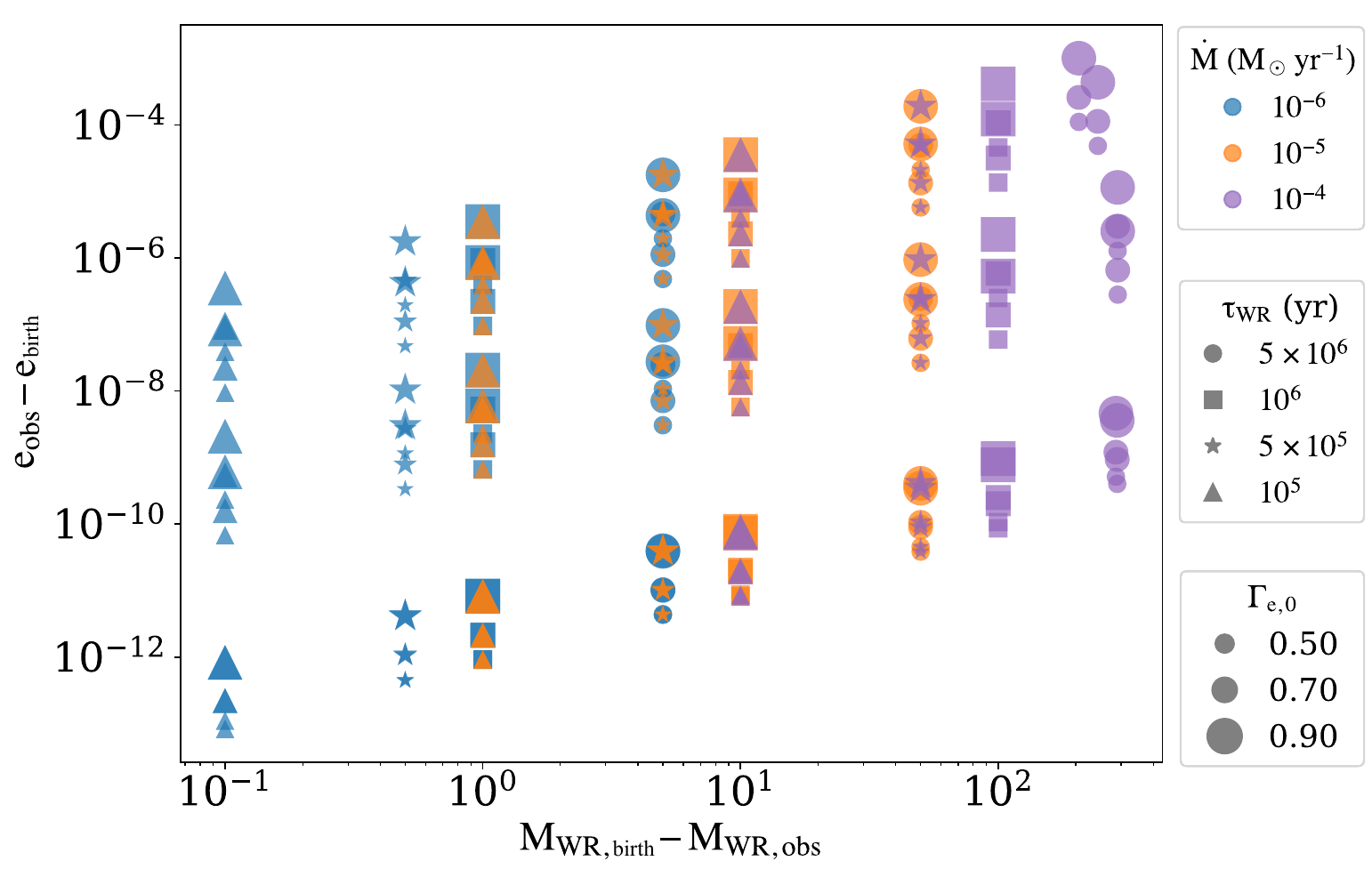}
    \caption{
    Eccentricity excitation as a function of total mass loss (both in terms of the difference between the currently observed value and the value at birth, $\tau_{\rm WR}$ time in the past) in the investigated WR binaries assuming tidally enhanced mass loss.
    Mass loss rates are represented by different colors, WR lifetimes by symbols, and Eddington parameters by symbol sizes.
    }
    \label{fig:tidal-massloss}
\end{figure}

The previous section identifies the systems whose reconstructed birth separations place them in the $\tau_{\rm tidal}\ll\tau_{\rm WR}$ regime, which means that a primordial eccentricity could not have survived.
An in-situ pumping mechanism is then required to reproduce the currently observed eccentricities.
We present how much eccentricity pumping can occur in the investigated WR systems due to a tidally enhanced mass loss that peaks at the pericenter of the orbit (see Sect.~\ref{sec:methods:orbit_averaging}, Eq.~\ref{eq:mdot_peri}).
We tested mass loss rates of $\dot{M}\in[10^{-6};10^{-5};10^{-4}]~M_\odot~\mathrm{yr^{-1}}$, WR lifetimes of $\tau_{\mathrm{WR}}\in[10^5;5\times10^5;10^6;5\times10^6]~\mathrm{yr}$ and Eddington parameters of $\Gamma_{e,0}\in[0.5;0.7;0.9]$.

Figure~\ref{fig:tidal-massloss} shows the change in eccentricity after backwards integration of the WR systems for $\tau_{\rm WR}$ time, investigated as a function of the mass lost by the WR star.
Different mass loss rates and WR lifetimes are denoted by colors and symbols, while symbol sizes show various $\Gamma_{e,0}$ values.
It is clear that for more significant total mass loss (corresponding to higher mass loss rates and longer WR lifetimes), the change in eccentricity is also more significant.
One can also see that a larger Eddington parameter also drives a more significant change in eccentricity.

The eccentricities shown in Fig.~\ref{fig:tidal-massloss} are the values that can be gained as a consequence of tidally enhanced mass loss alone.
If $\Delta e \ll e_\mathrm{obs}$ for a given system, then the tidal channel is incapable of explaining a significant fraction of the observed eccentricity, and most of $e_\mathrm{obs}$ must be attributed either to the primordial state inherited from the pre-interaction binary or to a different evolutionary mechanism acting during the WR phase.
Conversely, if $\Delta e \sim e_\mathrm{obs}$, the tidal channel could plausibly account for the observed eccentricity from a nearly circular primordial state.
As one can see, the change in eccentricity is $\Delta e\lesssim10^{-4}$ even for the most significant mass loss rates and longest WR lifetimes.
This is orders of magnitude below the eccentricity of all of the non-circularized models in our sample, which means that tidal enhancement of mass loss cannot account for the observed eccentricities of WR binaries.

The change in semi-major axis across the systems falls into a range of $a_{\rm obs}/a_{\rm birth}\in[1.0007;10.1]$.
This correlates well with the findings of the classical adiabatic mass loss model presented in Sect.~\ref{sec:res:adiabatic}.

For the longest lifetime
($\tau_{\rm WR} = 5\times10^6$\,yr) combined with the highest mass-loss rate
($\dot{M} = 10^{-4}\,M_\odot\,{\rm yr}^{-1}$), the backward reconstruction of
WR\,140, WR\,137, WR\,11, WR\,21a, WR\,22, and WR\,133 drives the WR mass at
birth up to the $300\,M_\odot$ ceiling.
In this case, these systems require a progenitor mass in the regime of the most massive known
stars ($\sim 250$--$300\,M_\odot$; \citealt{Crowther2010,
Bestenlehner2020}), where the objects are young, hydrogen-rich main-sequence
stars rather than evolved classical WR stars.
As such, we discard these runs, retaining them only as upper limits on the reconstructed
mass and eccentricity change.
The corresponding $\Delta e$ values for the discarded runs span $10^{-10}$ to $10^{-3}$ but are not used in our
analysis.
Crucially, this does not affect our conclusions.

Because the tidal channel supplies $\Delta e\lesssim10^{-4}$, it cannot be the required mechanism. 
For the systems flagged in Sect~\ref{sec:res:tidal-circ}, this leaves only two possibilities: 
either their eccentricity is in fact primordial (implying their true mass-loss rate lies at the low end of our range, so that they never entered the $\tau_{\rm tidal}\ll\tau_{\rm WR}$ regime) or an anisotropic channel we do not model here is responsible.

\section{Discussion and caveats}
\label{sec:discussion}

The principal result of our work is that tidally enhanced, pericenter-focused mass loss produces a secular eccentricity change of $\Delta e \lesssim 10^{-4}$ even under the most extreme combination of mass-loss rate ($\dot{M} = 10^{-4}, M_\odot,\mathrm{yr}^{-1}$), Eddington parameter ($\Gamma_{e,0} = 0.9$), and WR lifetime ($5\times10^{6}$yr) in our grid (Fig.\ref{fig:tidal-massloss}).
This is three to four orders of magnitude below the observed eccentricities of every non-circularized system in the sample. 
We therefore conclude that the eccentricities of WR binaries are largely primordial and inherited from the pre-interaction O+O configuration rather than generated during the WR phase by the tidal enhancement of the mass-loss rate. 

The two effectively circular systems in our sample with well-constrained component masses and separations, WR~20a and WR~139, lie close to the $\tau_{\mathrm{tidal}} = \tau_{\mathrm{WR}}$ boundary already at the present epoch, and our backward integration shows that both cross firmly into the $\tau_{\mathrm{tidal}} \ll \tau_{\mathrm{WR}}$ regime at their reconstructed birth separations. 
The prediction of the tidal framework is therefore unambiguous: 
any primordial eccentricity these systems may have possessed should have been damped to zero within the WR lifetime. 
Their observed near-circular orbits are exactly what this prediction requires. 

Regarding WR~11 and WR~133, it is possible for a mass loss rate of $\gtrsim 5\times10^{-5}~M_\odot~\rm yr^{-1}$ to circularize the orbit (see Fig.~\ref{fig:tidal-timescales}).
These systems, however, show moderate eccentricities of $\simeq 0.3$, which can not be reproduced by the tidal enhancement of the mass loss alone.
This suggests that the observed eccentricities may be primordial in origin; however, whether moderately anisotropic mass loss could also contribute to their excitation in these systems remains to be examined.
The same stands for WR~22 and WR~21a, where a mass loss rate of $\simeq 10^{-4}~M_\odot~\rm yr^{-1}$ is required for the systems to have been circularized during their lifetime.
As their eccentricities are 0.6 and 0.7, respectively, this means that their eccentricities are likely primordial assuming a plausible mass loss rate, or they require an even more heavily anisotropic mass loss geometry than WR~11 and WR~133.

For the widest systems, WR~140 and WR~137 ($e=0.9$ and $e = 0.3$, respectively, and $a\simeq15~\rm au$ for both), tidal circularisation can be basically ruled out over the complete mass loss rate range.
This is because the orbital separations ($\simeq16~\rm au$) are far too large for any continuous wind perturbation to accumulate a meaningful eccentricity change within the available time. 
For these systems, the circularization timescale exceeds the WR lifetime by many orders of magnitude (Fig.~\ref{fig:tidal-timescales}), so a primordial eccentricity is trivially preserved and requires no in-situ maintenance. 

Four systems in our sample (dust-producing binaries Apep, WR~112, WR~125, and WR~48a) lack published component masses and so cannot be integrated directly.
As an orbital period is available for each (see Tab.~\ref{tab:wr_binaries}), a conservative estimate of the semi-major axis is possible via Kepler's third law ($a = (M_{\rm tot}\,P^2)^{1/3}$) given an assumed total mass.
We adopt literature-based values of $M_{\rm tot} \approx 40\,M_\odot$ for WR~112 (\citealp{Lau2020}), $M_{\rm tot} \approx 35\,M_\odot$ for WR~125 \citep{richardson2024wr125}, $\approx 50\,M_\odot$ for WR 48a \citep{Williams2012}, and $M_{\rm tot} \approx 30\,M_\odot$ for Apep \citep{han2020apep, white2025apep}.
This yields $a \approx 25$, $30$, $37$ and $104\,{\rm au}$ for WR~112, WR~125, WR~48a and Apep, respectively.
All the above separations exceed those of every system in our integrable sample, placing the dusty systems firmly outside the regime in which tides operate.
Evaluating the circularization timescale of Eq.~(\ref{eq:ctl_tau_e_ps}) at these separations, with $R_{\rm WR} = 1\,R_\odot$ (as adopted elsewhere for radius-less systems), gives $\tau_e \sim 6\times10^{29}\,{\rm yr}$ for the tightest case (WR~112) and $\tau_e \sim 2\times10^{31}\,{\rm yr}$ for the widest (Apep).
These values are longer than the WR lifetime by 23-24 orders of magnitude.
Even if $R_{\rm WR} = 10\,R_\odot$, the timescales remain $\tau_e \gtrsim 6\times10^{24}\,{\rm yr}$, still some 18 orders of magnitude longer than any plausible WR phase.
As such, the corresponding tidally driven eccentricity change is negligible
and lies far below the $\Delta e \lesssim 10^{-4}$ bound established for the integrable sample.
We therefore conclude that any eccentricity in these wide dust systems is likely primordial.

The more interesting question for these wide systems is the origin of such extreme eccentricity in the first place, especially in the case of WR~140. 
We argue that the most likely channel is an impulsive, asymmetric mass-ejection event during a prior evolutionary phase, for instance, a luminous blue variable (LBV) -type eruption of the progenitor. 
Mass ejected anisotropically imparts a momentum kick to the orbit, which can introduce or amplify eccentricity (e.g., \citealp{Wang2021}, in the analogous context of shell flashes on proto-white dwarfs).
Repeated periastron-localized eruptions in eccentric massive binaries have been shown to drive orbital evolution directly \citep{Smith2011}. 
This is consistent with the observation that the highest-eccentricity WR systems are also among the widest, where impulsive perturbations survive undamped, whereas any close-orbit history would have been erased.

Furthermore, fossil ejecta from earlier eruptive phases are directly observed around a number of evolved massive stars.
The clearest example is the nebula M1-67 around WR~124, whose bipolar, CNO-processed structure has been attributed to an ejection during a LBV phase of a $\sim\!60\,M_\odot$ progenitor \citep{Fernandez-Martin2013}.
In addition, systematic mid-infrared surveys with WISE and Spitzer have revealed circumstellar shells around a substantial fraction of Galactic WR stars, tracing mass ejected in slow red supergiant or LBV winds and subsequently swept up by the fast WR wind \citep{Toala2015}.
Regarding our sample of systems, on the one hand, the wide dust producers (WR~140, WR~137, and Apep) are dominated by colliding-wind dust that traces the current orbital motion rather than fossil ejecta.
On the other hand, the WNha systems (WR~20a, WR~21a, WR~22, WR~25), being still core-hydrogen-burning \citep[e.g.][]{SmithConti2008}, are not expected to have undergone an LBV-type eruption, so for these objects a primordial origin of the eccentricity is favored over impulsive pumping.

The four WNha systems in our sample (WR~20a, WR~21a, WR~22, and WR~25) are very massive, near-Eddington objects that are core-hydrogen-burning and thus effectively main-sequence stars \citep{GrafenerHamann2008, Bestenlehner2020, sander2020}.
For example, spectral modelling of WR~22 places it at an age of only $\sim 2\,{\rm Myr}$ and implies that, even at its present mass-loss rate, it could have shed at most $\sim 28\,M_\odot$ over its lifetime, indicating that its current high-$\dot{M}$ phase is recent \citep{GrafenerHamann2008}.
Consequently, integrating these systems backward over the full canonical WR lifetime and mass-loss rates overestimates their cumulative mass loss and the associated orbital evolution.
As our reported eccentricity change is an upper limit obtained under the most extreme cases, overestimating the WR-phase duration of the WNh systems results in overestimating their tidally driven eccentricity pumping as well.
This means that the true $\Delta e$ for these younger objects is smaller still and the tidal channel remains unable to reproduce the observed eccentricities.
This also implies that the reconstructed birth separations of the WNh systems should be regarded as upper bounds.
Combined, these reinforce a primordial origin for the WNh systems' eccentricities.
We note that a fully self-consistent treatment would require a time-dependent, evolutionary-stage-specific mass-loss prescription that is beyond the scope of the present orbit-averaged calculation.

We refrain from quoting the quantitative mass-loss rate, or WR lifetime that would be required to reproduce the observed eccentricities through this channel, as our grid is already constructed to span -- and at its endpoints, slightly exceed -- the physically and observationally admissible range.
Thus, any value lying outside it describes an object that is no longer a Wolf–Rayet star (for instance, requiring sustained mass loss of order a solar mass per year over a Gyr).
Since the Eddington-pumping magnitude scales linearly with $k_2$ through $\varepsilon_{\rm T}$, a larger Love number would raise $\Delta e$ by at most an order of magnitude to $10^{-3}$ - still three to four orders of magnitude below the observed eccentricities.
The conclusion that tidal enhancement of mass loss cannot have driven the high eccentricities of WR binaries is therefore robust to the adopted value of $k_2$.

The failure of the tidally enhanced channel does not imply that wind-driven eccentricity pumping is impossible.
It does, however, require that the geometry of mass ejection be more sophisticated. 
In the formalism of \citet{DosopoulouKalogera2016a, DosopoulouKalogera2016b}, an isotropic wind preserves the rotational symmetry of the system and contributes zero secular eccentricity change, exactly as we recover in our isotropic limit.
A genuinely anisotropic outflow, however, removes linear momentum along a body-fixed axis and can achieve $\langle \dot{e} \rangle \neq 0$, independent of how strongly the rate is modulated. 
This also emerges in the hydrodynamical simulations of \citet{SaladinoPols2019}, who find that once the outflow departs from the fast-isotropic limit, the orbit-averaged eccentricity can actually grow.
A full treatment of this channel requires evolving the orbital elements under an anisotropic-momentum prescription within a variable-mass N-body framework, which lies beyond the scope of the present orbit-averaged calculation.
However, we identify it here as the most promising in-situ mechanism and defer its quantitative exploration to future work.

The findings presented here should be interpreted in the light of several important caveats.
First, we have held the companion mass $M_2$ fixed throughout the integration.
For the WR\,+\,O systems that constitute most of our sample, this is well controlled: 
normal O- and B-type companions lose mass through stellar winds at rates one to two orders of magnitude below those of the WR primary \citep{NugisLamers2000}, so the neglected companion mass loss enters at the few-percent level in the orbital elements, far below the three-to-four-order-of-magnitude margin by which the tidal channel fails to reproduce the observed eccentricities. 

The above justification does not hold, however, for systems in which the companion is itself a Wolf--Rayet star. 
The triple star system Apep hosts a classical WC8\,+\,WN4-6b inner binary with a bound O-supergiant tertiary at $\sim$1700\,AU \citep{Callingham2020, white2025apep}; WR\,48a shows evidence for a compact inner binary ($P \approx 1$\,yr) nested within its longer-period ($P > 23$\,yr) dust-producing orbit \citep{Williams2012}; and WR\,20a is a near-equal-mass WN6ha\,+\,WN6ha pair. 
In these systems both components shed mass at comparable, WR-level rates, so setting $\dot{M}_2 = 0$ (Section~\ref{sec:res:adiabatic}) is no longer a small correction.
The true adiabatic widening and the tidal enhancement would both be modified, and our reconstructed birth separations for these systems should be regarded as correspondingly more uncertain. 

We also have to note a further dynamical channel that our two-body treatment does not capture:
the influence of a bound third body, which is especially relevant given the high multiplicity of massive and WR stars.
Spectroscopic surveys find observed binary fractions of $\sim\!0.4$--$0.6$ across the Galactic WR sequences \citep{Dsilva2020, Dsilva2022, Dsilva2023}, and the true multiplicity is higher still once wider and hierarchical companions are accounted for.
Indeed, five systems in our own sample (Apep, WR~48a, WR~20a, WR~138 and WR~11) are known or suspected higher-order multiples.

Such a configuration opens two distinct eccentricity-generating pathways.
On the one hand, in a stable, sufficiently inclined hierarchical triple, the tertiary drives secular Kozai--Lidov oscillations that can raise the inner-binary eccentricity to $e_{\rm max} \approx \sqrt{1 - 5/3~\cos^2 (i)}$, approaching unity for near-perpendicular orbits, while leaving the inner semi-major axis essentially unchanged \citep{Kozai1962, Lidov1962, Naoz2016}.
On the other hand, if the triple is dynamically unstable, the system undergoes a chaotic interaction that typically ejects the lowest-mass component, modifying both $e$ and $a$.
Note that adiabatic WR mass loss can itself induce such a dynamical instability by widening the inner orbit toward the stability boundary and is thus an attractive explanation for the wide, highly eccentric systems.
A quantitative treatment of either pathway lies beyond the orbit-averaged, two-body framework adopted here; we therefore identify three-body dynamics as a promising mechanism for eccentricity growth of the multiple systems in our sample, and defer its quantitative exploration to future work.

Concerning the tidal circularization, we note that the pseudo-synchronization formula \citep{Hut1981} is derived in the weak-friction, constant-time-lag limit, and its accuracy degrades at high eccentricity. 
Detailed tidal-flow calculations confirm the Hut pseudo-synchronization angular velocity for moderate eccentricities ($e \lesssim 0.3$), but at the high-$e$ end of the sample (WR~140 at 0.90, WR~21a at 0.70) the precise equilibrium spin is more uncertain. 

We also need to note the eccentricity-generating mechanisms that we do not invoke, so that their relevance is clearly not understated.
We do not consider Kozai–Lidov oscillations driven by a bound third body: 
while several WR systems are known or suspected to belong to hierarchical triples, modeling that channel requires the (largely unconstrained) orbital architecture of the outer companion and is properly the subject of a dedicated study. 
Likewise, we do not model eccentricity imparted by an asymmetric supernova kick from a prior companion that is known to be a driver of eccentricity \citep{BrandtPodsiadlowski1995} but has the disadvantage of disrupting the stability of stellar systems (e.g., \citealp{regaly-etal-22, frohlich-etal-23}).
Our analysis is therefore conservative: 
it asks whether the secular wind physics intrinsic to the observed binary can account for the eccentricities, and concludes that it cannot through the rate channel.
This leaves primordial inheritance, impulsive progenitor-phase ejection, and anisotropic wind momentum as the surviving explanations.

All in all, our results favor that the eccentricities of WR binaries are predominantly primordial, inherited from the pre-interaction massive-binary population, with at most modest additional contributions from anisotropic wind momentum during the WR phase, and, for the most extreme systems, an impulsive, asymmetric ejection during a prior LBV-like phase \citep{Smith2011} as the probable origin of the eccentricity itself.
The comparison sample of Galactic visual and spectroscopic binaries (Fig.~\ref{fig:system-eccs}) reinforces this: 
massive ($\geq 20\, M_\odot$) systems show a clear preference for low eccentricity, with only two systems above $e \simeq 0.65$, so the high eccentricities of WR~140, Apep and WR~21a seem like genuine outliers relative to the field population and are unlikely to have been produced by any generic, continuously acting process.

\section{Conclusions}
\label{sec:conclusions}

In this study, we investigated whether the high orbital eccentricities observed in Wolf--Rayet binaries can be generated by a tidal enhancement of the mass-loss rate, or whether they must instead be primordial.
We investigated a sample of 14 WR binaries with measured orbital and stellar parameters and compared their eccentricity distribution with that of mass-matched samples of Galactic field binaries.
We modeled each system as a variable-mass two-body problem, integrating the secular orbital evolution backward in time over the WR lifetime under three successive prescriptions: 
i) adiabatic isotropic mass loss, ii) equilibrium-tide circularization, and iii) a tidally enhanced, pericenter-focused mass loss in which the companion-facing reduction of the effective surface gravity raises the local Eddington parameter.

The backward integration reconstructs the birth separation of each system, identifies those whose early orbits were tight enough for tides to have erased any primordial eccentricity, and quantifies how much eccentricity the tidally enhanced wind could pump over the available WR lifetime.
Our main findings are as follows.

\begin{enumerate}
\item The eccentricity distribution of WR binaries differs from that of mass-matched ($\geq 20\,M_\odot$) field binaries at high significance.
The discrepancy is driven by the high-eccentricity tail: only $\simeq 4\%$ of massive field binaries have $e \geq 0.6$, against $\simeq 29\%$ of the WR systems.
\item Adiabatic, isotropic mass loss widens the orbit while conserving eccentricity, with the semi-major axis growing by up to $a_{\rm obs}/a_{\rm birth}\simeq20$ for the least massive systems over a $5\times10^6$\,yr WR lifetime, and negligibly for the most massive ones.
\item Tidal circularization is efficient only for the tightest reconstructed birth separations ($\lesssim 0.5~\rm au$). 
Integrating backwards in time, reaching such compact orbits within the WR lifetime requires mass-loss rates at the upper end of the observed range ($\dot{M}\gtrsim5\times10^{-5}\,M_\odot\,{\rm yr}^{-1}$).
The near-circular systems WR\,20a and WR\,139 lie close to the $\tau_{\rm tidal}=\tau_{\rm WR}$ boundary already at the present epoch, consistent with their observed nearly circular orbits.
\item The tidally enhanced, pericenter-focused mass loss pumps the eccentricity by at most $10^{-4}$, even under the most extreme case of mass loss.
This change in eccentricity is orders of magnitude below the observed eccentricities of every non-circularized system.
This model therefore cannot reproduce the observed eccentricities.
Therefore, tidally circularizable systems face a dilemma: 
either their eccentricity is primordial (implying their mass-loss rate is so low that they never circularized) or a more elaborate mass-loss model is required to pump their eccentricities.
\item We conclude that WR eccentricities are predominantly primordial, inherited from the pre-interaction O+O configuration.
This is reinforced by the field-binary comparison, in which high eccentricity is rare among massive systems, marking WR\,140, Apep and WR\,21a as genuine outliers.
\end{enumerate}

Taken together, these results show that a negligible fraction of the observed WR eccentricity is attributable to the tidal modulation of the mass-loss rate.
The high eccentricities of WR binaries are thus not generated by the enhanced wind near pericenter.
On the one hand, they may be inherited from the primordial systems.
On the other hand, for the widest, moderate-to-high eccentricity systems (WR\,140, WR~137), eccentricity is pumped presumably by asymmetric mass-ejection events during a prior evolutionary phase, such as a luminous blue variable-type eruption of the progenitor, which is worth exploring in detail in a future study.
Finally, neither constant adiabatic WR winds nor time-variable, pericenter-enhanced winds are expected to substantially affect the future orbital evolution of the present WR systems; thus, in the absence of a third component, their current orbits can be regarded as stable.

\section*{Acknowledgements}
We thank the anonymous referee for their helpful comments, which significantly improved the quality of this paper.
Supported by the EKÖP-25-2-I-ELTE-358 University Research Scholarship Program of the Ministry for Culture and Innovation from the source of the National Research, Development and Innovation Fund. V. F. acknowledges financial support from the undergraduate research assistant program of the Konkoly Observatory.

\bibliography{WR}{}

@article{Williams1990,
  author  = {Williams, P. M. and van der Hucht, K. A. and Pollock, A. M. T. and Florkowski, D. R. and van der Woerd, H. and Wamsteker, W. M.},
  title   = {Multi-frequency variations of the Wolf-Rayet system HD 193793--I. Infrared, X-ray and radio observations},
  journal = {Monthly Notices of the Royal Astronomical Society},
  year    = {1990},
  volume  = {243},
  pages   = {662--684}
}

@article{Lieb2025_WR140,
  author       = {Lieb, Emma P. and Lau, Ryan M. and Hoffman, Jennifer L. and others},
  title        = {Dynamic Imprints of Colliding-wind Dust Formation from WR140},
  journal      = {The Astrophysical Journal Letters},
  year         = {2025},
  volume       = {979},
  number       = {1},
  pages        = {L3},
  doi          = {10.3847/2041-8213/ad9aa9},
  eprint       = {2502.02738},
  archivePrefix= {arXiv},
  primaryClass = {astro-ph.GA}
}

@INPROCEEDINGS{WhiteTuthill2024,
       author = {{White}, Ryan M.~T. and {Tuthill}, Peter},
        title = "{Wolf-Rayet colliding wind binaries}",
    booktitle = {Encyclopedia of Astrophysics, Volume 2},
         year = 2026,
       volume = {2},
        month = jan,
        pages = {584-603},
          doi = {10.1016/B978-0-443-21439-4.00067-5},
archivePrefix = {arXiv},
       eprint = {2412.12534},
 primaryClass = {astro-ph.SR},
       adsurl = {https://ui.adsabs.harvard.edu/abs/2026enap....2..584W}
}

@article{Hadjidemetriou1963,
  author  = {Hadjidemetriou, John D.},
  title   = {Two-Body Problem with Variable Mass: A New Approach},
  journal = {Icarus},
  year    = {1963},
  volume  = {2},
  pages   = {440--451},
  doi     = {10.1016/0019-1035(63)90072-1}
}

@article{DosopoulouKalogera2016b,
  author        = {Dosopoulou, Fani and Kalogera, Vicky},
  title         = {Orbital Evolution of Mass-transferring Eccentric Binary Systems. II. Secular Evolution},
  journal       = {The Astrophysical Journal},
  year          = {2016},
  volume        = {825},
  number        = {1},
  pages         = {71},
  doi           = {10.3847/0004-637X/825/1/71},
  archivePrefix = {arXiv},
  eprint        = {1603.06593},
  primaryClass  = {astro-ph.SR}
}

@article{SaladinoPols2019,
  author        = {Saladino, Martha I. and Pols, Onno R.},
  title         = {The eccentric behaviour of windy binary stars},
  journal       = {Astronomy \& Astrophysics},
  year          = {2019},
  volume        = {629},
  pages         = {A103},
  doi           = {10.1051/0004-6361/201935625},
  archivePrefix = {arXiv},
  eprint        = {1906.02038},
  primaryClass  = {astro-ph.SR}
}

@ARTICLE{RichardsonEtAl2024_WR137,
       author = {{Richardson}, Noel D. and {Schaefer}, Gail H. and {Eldridge}, Jan J. and {Spejcher}, Rebecca and {Holdsworth}, Amanda and {Lau}, Ryan M. and {Monnier}, John D. and {Moffat}, Anthony F.~J. and {Weigelt}, Gerd and {Williams}, Peredur M. and {Kraus}, Stefan and {Le Bouquin}, Jean-Baptiste and {Anugu}, Narsireddy and {Chhabra}, Sorabh and {Codron}, Isabelle and {Ennis}, Jacob and {Gardner}, Tyler and {Gutierrez}, Mayra and {Ibrahim}, Noura and {Labdon}, Aaron and {Lanthermann}, Cyprien and {Setterholm}, Benjamin R.},
        title = "{Visual Orbits of Wolf─Rayet Stars. I. The Orbit of the Dust-producing Wolf─Rayet Binary WR 137 Measured with the CHARA Array}",
      journal = {\apj},
         year = 2024,
        month = dec,
       volume = {977},
       number = {1},
          eid = {78},
        pages = {78},
          doi = {10.3847/1538-4357/ad8d5c},
archivePrefix = {arXiv},
       eprint = {2410.09259},
 primaryClass = {astro-ph.SR},
       adsurl = {https://ui.adsabs.harvard.edu/abs/2024ApJ...977...78R}
}

@article{crowther2007,
  author  = {Crowther, Paul A.},
  title   = {Physical Properties of {Wolf-Rayet} Stars},
  journal = {Annual Review of Astronomy and Astrophysics},
  year    = {2007},
  volume  = {45},
  pages   = {177--219},
  doi     = {10.1146/annurev.astro.45.051806.110615},
  eprint  = {astro-ph/0610356},
  archivePrefix = {arXiv}
}

@article{sander2020,
  author  = {Sander, Andreas A. C. and Vink, Jorick S. and Hamann, Wolf-Rainer},
  title   = {Driving classical {Wolf-Rayet} winds: a {$\Gamma$}- and {Z}-dependent mass-loss},
  journal = {Monthly Notices of the Royal Astronomical Society},
  year    = {2020},
  volume  = {491},
  number  = {3},
  pages   = {4406--4425},
  doi     = {10.1093/mnras/stz3064},
  eprint  = {1910.12886},
  archivePrefix = {arXiv},
  primaryClass  = {astro-ph.SR}
}

@article{lau2022,
  author  = {Lau, Ryan M. and Hankins, Matthew J. and Han, Yinuo and Argyriou, Ioannis and Corcoran, Michael F. and Eldridge, Jan J. and Endo, Izumi and Fox, Ori D. and Garcia Marin, Macarena and Gull, Theodore R. and Jones, Olivia C. and Hamaguchi, Kenji and Lamberts, Astrid and Law, David R. and Madura, Thomas and Marchenko, Sergey V. and Matsuhara, Hideo and Moffat, Anthony F. J. and Morris, Mark R. and Morris, Patrick W. and Onaka, Takashi and Ressler, Michael E. and Richardson, Noel D. and Russell, Christopher M. P. and Sanchez-Bermudez, Joel and Smith, Nathan and Soulain, Anthony and Stevens, Ian R. and Tuthill, Peter and Weigelt, Gerd and Williams, Peredur M. and Yamaguchi, Ryodai},
  title   = {Nested dust shells around the {Wolf-Rayet} binary {WR}~140 observed with {JWST}},
  journal = {Nature Astronomy},
  year    = {2022},
  volume  = {6},
  pages   = {1308--1316},
  doi     = {10.1038/s41550-022-01812-x},
  eprint  = {2210.06452},
  archivePrefix = {arXiv},
  primaryClass  = {astro-ph.SR}
}

@article{Hut1981,
  author  = {Hut, P.},
  title   = {Tidal evolution in close binary systems},
  journal = {Astronomy and Astrophysics},
  year    = {1981},
  volume  = {99},
  pages   = {126--140},
  adsurl  = {https://ui.adsabs.harvard.edu/abs/1981A%26A....99..126H/abstract}
}

@article{Veras2011,
  author  = {Veras, Dimitri and Wyatt, Mark C. and Mustill, Alexander J. and Bonsor, Amy and Eldridge, John J.},
  title   = {The great escape: how exoplanets and smaller bodies desert dying stars},
  journal = {Monthly Notices of the Royal Astronomical Society},
  year    = {2011},
  volume  = {417},
  number  = {3},
  pages   = {2104--2123},
  doi     = {10.1111/j.1365-2966.2011.19393.x}
}

@article{NugisLamers2000,
  author  = {Nugis, T. and Lamers, H. J. G. L. M.},
  title   = {Mass-loss rates of Wolf-Rayet stars as a function of stellar parameters},
  journal = {Astronomy \& Astrophysics},
  year    = {2000},
  volume  = {360},
  pages   = {227--244},
  adsbibcode = {2000A\&A...360..227N}
}

@article{EggletonKiselevaHut1998,
  author  = {Eggleton, P. P. and Kiseleva, L. G. and Hut, Piet},
  title   = {The equilibrium tide model for tidal friction},
  journal = {The Astrophysical Journal},
  year    = {1998},
  volume  = {499},
  pages   = {853--870},
  doi     = {10.1086/305670},
  eprint  = {astro-ph/9801246},
  archivePrefix = {arXiv}
}

@article{Ogilvie2014,
  author  = {Ogilvie, Gordon I.},
  title   = {Tidal Dissipation in Stars and Giant Planets},
  journal = {Annual Review of Astronomy and Astrophysics},
  year    = {2014},
  volume  = {52},
  number  = {1},
  pages   = {171--210},
  doi     = {10.1146/annurev-astro-081913-035941},
  eprint  = {1406.2207},
  archivePrefix = {arXiv},
  primaryClass  = {astro-ph.SR}
}

@article{PatelPenevSchussler2023,
  author  = {Patel, Shreyas S. and Penev, Kaloyan and Sch{\"u}ssler, Florian},
  title   = {Constraints on tidal quality factor in {Kepler} eclipsing binaries using tidal synchronization: a frequency-dependent approach},
  journal = {Monthly Notices of the Royal Astronomical Society},
  year    = {2023},
  volume  = {524},
  number  = {4},
  pages   = {5575--5590},
  doi     = {10.1093/mnras/stad2194}
}

@article{Crawford2025,
  author  = {Hobson-Ritz, Marshall and Birky, Jessica and Peterson, Leah and Gwartney, Peter and Wong, Rachel and Delker, John and Gordon, Tyler and Gilbert, Samantha and Davenport, James R. A. and Barnes, Rory},
  title   = {Tidal Synchronization of {TESS} Eclipsing Binaries},
  journal = {The Astrophysical Journal},
  year    = {2025},
  volume  = {990},
  number  = {2},
  pages   = {124},
  doi     = {10.3847/1538-4357/adf10d},
  eprint  = {2501.04082},
  archivePrefix = {arXiv},
  primaryClass  = {astro-ph.SR}
}

@article{Fleming2019,
  author  = {Fleming, David P. and Barnes, Rory and Davenport, James R. A. and Luger, Rodrigo},
  title   = {Rotation Period Evolution in Low-Mass Binary Stars: The Impact of Tidal Torques and Magnetic Braking},
  journal = {The Astrophysical Journal},
  year    = {2019},
  volume  = {881},
  pages   = {88},
  doi     = {10.3847/1538-4357/ab2ed2},
  eprint  = {1903.05686},
  archivePrefix = {arXiv},
  primaryClass  = {astro-ph.SR}
}

@article{Lurie2017,
  author  = {Lurie, John C. and Vyhmeister, Karl and Hawley, Suzanne L. and Adilia, Jamel and Chen, Andrea and Davenport, James R. A. and Juric, Mario and Puig-Holzman, Michael and Weisenburger, Kolby L.},
  title   = {Tidal Synchronization and Differential Rotation of {Kepler} Eclipsing Binaries},
  journal = {The Astronomical Journal},
  year    = {2017},
  volume  = {154},
  pages   = {250},
  doi     = {10.3847/1538-3881/aa974d},
  eprint  = {1710.07339},
  archivePrefix = {arXiv},
  primaryClass  = {astro-ph.SR}
}

@article{Claret2023,
  author  = {Claret, A.},
  title   = {Theoretical tidal evolution constants for stellar models from the pre-main sequence to the white dwarf stage: Apsidal motion constants, moment of inertia, and gravitational potential energy},
  journal = {Astronomy \& Astrophysics},
  year    = {2023},
  volume  = {674},
  pages   = {A67},
  doi     = {10.1051/0004-6361/202346250},
  eprint  = {2305.01627},
  archivePrefix = {arXiv},
  primaryClass  = {astro-ph.SR}
}

@article{ClaretGimenezEtAl2021,
  author  = {Claret, A. and Gim{\'e}nez, A. and Baroch, D. and Ribas, I. and Morales, J. C. and Anglada-Escud{\'e}, G.},
  title   = {Analysis of apsidal motion in eclipsing binaries using {TESS} data. {II}. A test of internal stellar structure},
  journal = {Astronomy \& Astrophysics},
  year    = {2021},
  volume  = {654},
  pages   = {A17},
  doi     = {10.1051/0004-6361/202141484},
  eprint  = {2107.10765},
  archivePrefix = {arXiv},
  primaryClass  = {astro-ph.SR}
}

@article{thomas2021wr140,
  author  = {Thomas, J. D. and Monnier, J. D. and Richardson, N. D. and others},
  title   = {An orbit and dynamical masses for the colliding-wind binary WR 140 from long-baseline interferometry and radial velocities},
  journal = {Monthly Notices of the Royal Astronomical Society},
  year    = {2021},
  volume  = {504},
  pages   = {5221--5237},
  doi     = {10.1093/mnras/stab1117}
}

@ARTICLE{white2025apep,
       author = {{White}, Ryan M.~T. and {Pope}, Benjamin J.~S. and {Tuthill}, Peter G. and {Han}, Yinuo and {Dholakia}, Shashank and {Lau}, Ryan M. and {Callingham}, Joseph R. and {Richardson}, Noel D.},
        title = "{The Serpent Eating Its Own Tail: Dust Destruction in the Apep Colliding Wind Nebula}",
      journal = {\apj},
         year = 2025,
        month = nov,
       volume = {994},
       number = {1},
          eid = {121},
        pages = {121},
          doi = {10.3847/1538-4357/adfbe1},
archivePrefix = {arXiv},
       eprint = {2507.14610},
 primaryClass = {astro-ph.SR},
       adsurl = {https://ui.adsabs.harvard.edu/abs/2025ApJ...994..121W}
}

@ARTICLE{richardson2024wr125,
       author = {{Richardson}, Noel D. and {Daly}, Andrea R. and {Williams}, Peredur M. and {Hill}, Grant M. and {Shenavrin}, Victor I. and {Endo}, Izumi and {Chen{\'e}}, Andr{\'e}-Nicolas and {Karnath}, Nicole and {Lau}, Ryan M. and {Moffat}, Anthony F.~J. and {Weigelt}, Gerd},
        title = "{The Long-period Spectroscopic Orbit and Dust Creation in the Wolf─Rayet Binary System WR 125}",
      journal = {\apj},
         year = 2024,
        month = jul,
       volume = {969},
       number = {2},
          eid = {140},
        pages = {140},
          doi = {10.3847/1538-4357/ad4d54},
archivePrefix = {arXiv},
       eprint = {2405.10454},
 primaryClass = {astro-ph.SR},
       adsurl = {https://ui.adsabs.harvard.edu/abs/2024ApJ...969..140R}
}

@ARTICLE{shaposhnikov2023v444cyg,
       author = {{Shaposhnikov}, I. and {Cherepashchuk}, A. and {Dodin}, A. and {Postnov}, K.},
        title = "{Spectroscopic searches for evolutionary orbital period changes in WR+OB binaries: the case of V444 Cyg}",
      journal = {\mnras},
         year = 2023,
        month = dec,
       volume = {526},
       number = {3},
        pages = {4529-4534},
          doi = {10.1093/mnras/stad2859},
archivePrefix = {arXiv},
       eprint = {2309.08386},
 primaryClass = {astro-ph.SR},
       adsurl = {https://ui.adsabs.harvard.edu/abs/2023MNRAS.526.4529S}
}

@article{north2007gamma2vel,
  author  = {North, J. R. and Tuthill, P. G. and Tango, W. J. and Davis, J.},
  title   = {$\gamma^2$ Velorum: orbital solution and fundamental parameter determination with SUSI},
  journal = {Monthly Notices of the Royal Astronomical Society},
  year    = {2007},
  volume  = {377},
  number  = {1},
  pages   = {415--424},
  doi     = {10.1111/j.1365-2966.2007.11608.x}
}

@ARTICLE{williamsperedur11,
       author = {{Williams}, Peredur},
        title = "{Results from the 2009 campaign on WR 140}",
      journal = {Bulletin de la Societe Royale des Sciences de Liege},
         year = 2011,
        month = jan,
       volume = {80},
        pages = {595-609},
       adsurl = {https://ui.adsabs.harvard.edu/abs/2011BSRSL..80..595W}
}

@ARTICLE{schnurreetal09,
       author = {{Schnurr}, O. and {Moffat}, A.~F.~J. and {Villar-Sbaffi}, A. and {St-Louis}, N. and {Morrell}, N.~I.},
        title = "{A first orbital solution for the very massive 30 Dor main-sequence WN6h+O binary R145}",
      journal = {\mnras},
         year = 2009,
        month = may,
       volume = {395},
       number = {2},
        pages = {823-836},
          doi = {10.1111/j.1365-2966.2009.14437.x},
archivePrefix = {arXiv},
       eprint = {0901.0698},
 primaryClass = {astro-ph.SR},
       adsurl = {https://ui.adsabs.harvard.edu/abs/2009MNRAS.395..823S}
}

@ARTICLE{rauwetal04,
       author = {{Rauw}, G. and {Crowther}, P.~A. and {De Becker}, M. and {Gosset}, E. and {Naz{\'e}}, Y. and {Sana}, H. and {van der Hucht}, K.~A. and {Vreux}, J.-M. and {Williams}, P.~M.},
        title = "{The spectrum of the very massive binary system WR 20a (WN6ha + WN6ha): Fundamental parameters and wind interactions}",
      journal = {\aap},
         year = 2005,
        month = mar,
       volume = {432},
       number = {3},
        pages = {985-998},
          doi = {10.1051/0004-6361:20042136},
       adsurl = {https://ui.adsabs.harvard.edu/abs/2005A&A...432..985R}
}

@ARTICLE{schmutzetal26,
       author = {{Schmutz}, Werner and {Hummel}, Christian A. and {Koenigsberger}, Gloria and {Millour}, Florentin and {Sanchez-Bermudez}, Joel},
        title = "{Revised orbital parameters of the gamma2 Velorum system}",
      journal = {arXiv e-prints},
         year = 2026,
        month = jun,
          eid = {arXiv:2606.27265},
        pages = {arXiv:2606.27265},
          doi = {10.48550/arXiv.2606.27265},
archivePrefix = {arXiv},
       eprint = {2606.27265},
 primaryClass = {astro-ph.SR},
       adsurl = {https://ui.adsabs.harvard.edu/abs/2026arXiv260627265S}
}

@ARTICLE{barbaetal22,
       author = {{Barb{\'a}}, Rodolfo H. and {Gamen}, Roberto C. and {Mart{\'\i}n-Ravelo}, Pablo and {Arias}, Julia I. and {Morrell}, Nidia I.},
        title = "{The winking eye of a very massive star: WR 21a revealed as an eclipsing binary by TESS}",
      journal = {\mnras},
         year = 2022,
        month = oct,
       volume = {516},
       number = {1},
        pages = {1149-1157},
          doi = {10.1093/mnras/stac2173},
archivePrefix = {arXiv},
       eprint = {2109.06311},
 primaryClass = {astro-ph.SR},
       adsurl = {https://ui.adsabs.harvard.edu/abs/2022MNRAS.516.1149B}
}

@INPROCEEDINGS{richardsonetal21,
       author = {{Richardson}, N.~D. and {Schaefer}, G. and {Thomas}, J. and {Lee}, L. and {Eldridge}, J.~J. and {Sander}, A. and {Shenar}, T.},
        title = "{Visual Orbits for Wolf-Rayet stars from CHARA Interferometry}",
    booktitle = {American Astronomical Society Meeting Abstracts \#238},
         year = 2021,
       series = {American Astronomical Society Meeting Abstracts},
       volume = {238},
        month = jun,
          eid = {308.07},
        pages = {308.07},
       adsurl = {https://ui.adsabs.harvard.edu/abs/2021AAS...23830807R}
}

@ARTICLE{Hamannetal2019,
       author = {{Hamann}, W.-R. and {Gr{\"a}fener}, G. and {Liermann}, A. and {Hainich}, R. and {Sander}, A.~A.~C. and {Shenar}, T. and {Ramachandran}, V. and {Todt}, H. and {Oskinova}, L.~M.},
        title = "{The Galactic WN stars revisited. Impact of Gaia distances on fundamental stellar parameters}",
      journal = {\aap},
         year = 2019,
        month = may,
       volume = {625},
          eid = {A57},
        pages = {A57},
          doi = {10.1051/0004-6361/201834850},
archivePrefix = {arXiv},
       eprint = {1904.04687},
 primaryClass = {astro-ph.SR},
       adsurl = {https://ui.adsabs.harvard.edu/abs/2019A&A...625A..57H}
}

@ARTICLE{gvaramadzeetal09,
       author = {{Gvaramadze}, V.~V. and {Fabrika}, S. and {Hamann}, W.-R. and {Sholukhova}, O. and {Valeev}, A.~F. and {Goranskij}, V.~P. and {Cherepashchuk}, A.~M. and {Bomans}, D.~J. and {Oskinova}, L.~M.},
        title = "{Discovery of a new Wolf-Rayet star and its ring nebula in Cygnus}",
      journal = {\mnras},
         year = 2009,
        month = nov,
       volume = {400},
       number = {1},
        pages = {524-530},
          doi = {10.1111/j.1365-2966.2009.15492.x},
archivePrefix = {arXiv},
       eprint = {0909.0621},
 primaryClass = {astro-ph.SR},
       adsurl = {https://ui.adsabs.harvard.edu/abs/2009MNRAS.400..524G}
}

@article{tramper2016wr21a,
  author  = {Tramper, F. and Sana, H. and Fitzsimons, N. E. and de Koter, A. and Kaper, L. and Mahy, L. and Moffat, A.},
  title   = {The mass of the very massive binary WR21a},
  journal = {Monthly Notices of the Royal Astronomical Society},
  year    = {2016},
  volume  = {455},
  number  = {2},
  pages   = {1275--1281},
  doi     = {10.1093/mnras/stv2373},
  eprint  = {1510.03609},
  archivePrefix = {arXiv},
  primaryClass  = {astro-ph.SR}
}

@article{gosset2009wr22,
  author  = {Gosset, E. and Naz{\'e}, Y. and Sana, H. and Rauw, G.},
  title   = {Phase-resolved XMM-Newton observations of the massive WR+O binary WR 22},
  journal = {Astronomy \& Astrophysics},
  year    = {2009},
  volume  = {508},
  pages   = {805--821}
}

@article{gamen2006wr25,
  author  = {Gamen, R. and Gosset, E. and Morrell, N. and Niemela, V. and Sana, H. and Naz{\'e}, Y. and Rauw, G. and Barb{\'a}, R. and Solivella, G.},
  title   = {The first orbital solution for the massive colliding-wind binary HD 93162 ($\equiv$ WR 25)},
  journal = {Astronomy \& Astrophysics},
  year    = {2006},
  volume  = {460},
  pages   = {777--782},
  doi     = {10.1051/0004-6361:20065618},
  eprint  = {astro-ph/0609454},
  archivePrefix = {arXiv}
}

@ARTICLE{Pourbaix2004,
  author  = {Pourbaix, D. and Tokovinin, A. A. and Batten, A. H. and Fekel, F. C. and Hartkopf, W. I. and Levato, H. and Morrell, N. I. and Torres, G. and Udry, S.},
  title   = {SB9: The ninth catalogue of spectroscopic binary orbits},
  journal = {A\&A},
  year    = {2004},
  volume  = {424},
  pages   = {727--732},
  doi     = {10.1051/0004-6361:20041213}
}

@MISC{SBORBITCAT,
  author       = {{HEASARC}},
  title        = {SBORBITCAT: Spectroscopic Binary Orbits Ninth Catalog (Dynamic Version)},
  howpublished = {\url{https://heasarc.gsfc.nasa.gov/W3Browse/all/sborbitcat.html}},
  note         = {HEASARC dynamic mirror of SB9, accessed 2026-03-24}
}

@ARTICLE{GaiaDR3Summary,
  author  = {{Gaia Collaboration}},
  title   = {Gaia Data Release 3: Summary of the content and survey properties},
  journal = {A\&A},
  year    = {2023},
  volume  = {674},
  pages   = {A1},
  doi     = {10.1051/0004-6361/202243940}
}

@ARTICLE{Hartkopf2001,
       author = {{Hartkopf}, William I. and {Mason}, Brian D. and {Worley}, Charles E.},
        title = "{The 2001 US Naval Observatory Double Star CD-ROM. II. The Fifth Catalog of Orbits of Visual Binary Stars}",
      journal = {\aj},
         year = 2001,
       volume = {122},
       number = {6},
        pages = {3472-3479},
          doi = {10.1086/323921},
       adsurl = {https://ui.adsabs.harvard.edu/abs/2001AJ....122.3472H},
}

@ARTICLE{Halbwachs2023,
       author = {{Halbwachs}, Jean-Louis and {Pourbaix}, Dimitri and {Arenou}, Fr{\'e}d{\'e}ric and {Galluccio}, Laurent and {Guillout}, Patrick and {Bauchet}, Nathalie and {Marchal}, Olivier and {Sadowski}, Gilles and {Teyssier}, David},
        title = "{Gaia Data Release 3. Astrometric binary star processing}",
      journal = {\aap},
         year = 2023,
       volume = {674},
          eid = {A9},
        pages = {A9},
          doi = {10.1051/0004-6361/202243969},
       eprint = {2206.05726},
 primaryClass = {astro-ph.SR},
       adsurl = {https://ui.adsabs.harvard.edu/abs/2023A&A...674A...9H},
}

@ARTICLE{GaiaDR3NSS,
       author = {{Gaia Collaboration} and {Arenou}, F. and {Babusiaux}, C. and {Barstow}, M.~A. and {Faigler}, S. and {Jorissen}, A. and {Kervella}, P. and {Mazeh}, T. and {Mowlavi}, N. and {Panuzzo}, P. and {Sahlmann}, J. and {Shahaf}, S. and {S{\o}rensen}, M. and {Teyssier}, D. and others},
        title = "{Gaia Data Release 3. Stellar multiplicity, a teaser for the hidden treasure}",
      journal = {\aap},
         year = 2023,
       volume = {674},
          eid = {A34},
        pages = {A34},
          doi = {10.1051/0004-6361/202243782},
       eprint = {2206.05595},
 primaryClass = {astro-ph.SR},
       adsurl = {https://ui.adsabs.harvard.edu/abs/2023A&A...674A..34G},
}

@MISC{ORB6,
       author = {{Matson}, Rachel A. and {Williams}, Samuel J. and {Hartkopf}, William I. and {Mason}, Brian D.},
        title = {Sixth Catalog of Orbits of Visual Binary Stars},
 howpublished = {U.S. Naval Observatory, Washington, DC; \url{https://www.astro.gsu.edu/wds/orb6.html}},
         year = 2001,
         note = {Continuously updated; accessed 2026-03-24},
}

@book{Kopal1959,
  author    = {Kopal, Zden{\v{e}}k},
  title     = {Close Binary Systems},
  series    = {The International Astrophysics Series},
  volume    = {5},
  publisher = {Chapman \& Hall},
  address   = {London},
  year      = {1959},
  adsurl    = {https://ui.adsabs.harvard.edu/abs/1959cbs..book.....K},
  bibcode   = {1959cbs..book.....K}
}

@book{Love1911,
  author    = {Love, Augustus Edward Hough},
  title     = {Some Problems of Geodynamics},
  publisher = {Cambridge University Press},
  address   = {Cambridge},
  year      = {1911},
  adsurl    = {https://ui.adsabs.harvard.edu/abs/1911spge.book.....L},
  bibcode   = {1911spge.book.....L}
}

@article{FriendAbbott1986,
    author  = {Friend, D.~B. and Abbott, D.~C.},
    title   = {The Theory of Radiatively Driven Stellar Winds. {III}. Wind Models with Finite Disk Correction and Rotation},
    journal = {\apj},
    year    = {1986},
    volume  = {311},
    pages   = {701--710},
    doi     = {10.1086/164808}
}

@article{SanderVinkHamann2020,
    author  = {Sander, A.~A.~C. and Vink, J.~S. and Hamann, W.-R.},
    title   = {Driving classical {Wolf--Rayet} winds: a {$\Gamma$}- and {Z}-dependent mass-loss},
    journal = {\mnras},
    year    = {2020},
    volume  = {491},
    number  = {3},
    pages   = {4406--4425},
    doi     = {10.1093/mnras/stz3064},
    eprint  = {1910.12886},
    archiveprefix = {arXiv}
}

@article{Bestenlehner2020,
    author  = {Bestenlehner, J.~M.},
    title   = {Mass loss and the {Eddington} parameter: a new mass-loss recipe for hot and massive stars},
    journal = {\mnras},
    year    = {2020},
    volume  = {493},
    number  = {3},
    pages   = {3938--3951},
    doi     = {10.1093/mnras/staa474},
    eprint  = {2002.05168},
    archiveprefix = {arXiv}
}

@article{MorenoKoenigsbergerHarrington2011,
    author  = {Moreno, E. and Koenigsberger, G. and Harrington, D.~M.},
    title   = {Eccentric binaries: Tidal flows and periastron events},
    journal = {\aap},
    year    = {2011},
    volume  = {528},
    pages   = {A48},
    doi     = {10.1051/0004-6361/201015874},
    eprint  = {1102.4301},
    archiveprefix = {arXiv}
}

@article{KoenigsbergerMoreno2009,
    author  = {Koenigsberger, G. and Moreno, E.},
    title   = {Tidal effects and periastron events in binary stars},
    journal = {Revista Mexicana de Astronom{\'\i}a y Astrof{\'\i}sica Conference Series},
    year    = {2009},
    volume  = {36},
    pages   = {201--206},
    eprint  = {0903.1221},
    archiveprefix = {arXiv}
}

@article{KoenigsbergerEstrellaTrujillo2024,
    author  = {Koenigsberger, G. and Estrella-Trujillo, D.},
    title   = {Eccentric binaries: Periastron events and tidal heating},
    journal = {\aap},
    year    = {2024},
    volume  = {686},
    pages   = {A127},
    doi     = {10.1051/0004-6361/202349075},
    eprint  = {2404.08774},
    archiveprefix = {arXiv}
}

@article{Fuller2017,
    author  = {Fuller, Jim},
    title   = {Heartbeat stars, tidally excited oscillations and resonance locking},
    journal = {\mnras},
    year    = {2017},
    volume  = {472},
    number  = {2},
    pages   = {1538--1564},
    doi     = {10.1093/mnras/stx2135},
    eprint  = {1706.05054},
    archiveprefix = {arXiv}
}

@article{Pabloetal2017,
    author  = {Pablo, H. and Richardson, N.~D. and Fuller, J. and Rowe, J. and Moffat, A.~F.~J. and Kuschnig, R. and Popowicz, A. and Handler, G. and Neiner, C. and Pigulski, A. and Wade, G.~A. and Weiss, W. and Buysschaert, B. and Ramiaramanantsoa, T. and Bratcher, A.~D. and others},
    title   = {Allow for {TEOs} in the massive {O}-star binary {$\iota$~Orionis}: a {BRITE} case study},
    journal = {\mnras},
    year    = {2017},
    volume  = {467},
    number  = {2},
    pages   = {2494--2503},
    doi     = {10.1093/mnras/stx207},
    eprint  = {1703.02086},
    archiveprefix = {arXiv}
}

@article{StevensPollock1994,
    author  = {Stevens, I.~R. and Pollock, A.~M.~T.},
    title   = {Stagnation-point flow in colliding-wind binary systems},
    journal = {\mnras},
    year    = {1994},
    volume  = {269},
    pages   = {226--234},
    doi     = {10.1093/mnras/269.2.226}
}

@article{GayleyOwockiCranmer1997,
    author  = {Gayley, K.~G. and Owocki, S.~P. and Cranmer, S.~R.},
    title   = {Sudden Radiative Braking in Colliding Hot-Star Winds},
    journal = {\apj},
    year    = {1997},
    volume  = {475},
    number  = {2},
    pages   = {786--797},
    doi     = {10.1086/303573}
}

@article{GrafenerHamann2008,
    author  = {Gr{\"a}fener, G. and Hamann, W.-R.},
    title   = {Mass loss from late-type {WN} stars and its
               {Z}-dependence: Very massive stars approaching the
               {Eddington} limit},
    journal = {\aap},
    year    = {2008},
    volume  = {482},
    pages   = {945--960},
    doi     = {10.1051/0004-6361:20066176}
}

@article{Grafeneretal2011,
    author  = {Gr{\"a}fener, G. and Vink, J.~S. and de~Koter, A.
               and Langer, N.},
    title   = {The {Eddington} factor as the key to understand the
               winds of the most massive stars: Evidence for a
               {$\Gamma$}-dependent wind performance number},
    journal = {\aap},
    year    = {2011},
    volume  = {535},
    pages   = {A56},
    doi     = {10.1051/0004-6361/201116701}
}

@article{Vinketal2011,
    author  = {Vink, J.~S. and Muijres, L.~E. and Anthonisse, B.
               and de~Koter, A. and Gr{\"a}fener, G. and Langer, N.},
    title   = {Wind modelling of very massive stars up to 300 solar
               masses},
    journal = {\aap},
    year    = {2011},
    volume  = {531},
    pages   = {A132},
    doi     = {10.1051/0004-6361/201116614}
}

@article{DosopoulouKalogera2016a,
    author  = {Dosopoulou, F. and Kalogera, V.},
    title   = {Orbital Evolution of Mass-Transferring Eccentric
               Binary Systems. {I}. Phase-Dependent Evolution},
    journal = {\apj},
    year    = {2016},
    volume  = {825},
    pages   = {70},
    doi     = {10.3847/0004-637X/825/1/70},
    eprint  = {1603.06592},
    archiveprefix = {arXiv}
}

@article{Sepinskyetal2007,
    author  = {Sepinsky, J.~F. and Willems, B. and Kalogera, V.
               and Rasio, F.~A.},
    title   = {Interacting Binaries with Eccentric Orbits: Secular
               Orbital Evolution Due to Conservative Mass Transfer},
    journal = {\apj},
    year    = {2007},
    volume  = {667},
    pages   = {1170--1184},
    doi     = {10.1086/520911}
}

@article{Sepinskyetal2009,
    author  = {Sepinsky, J.~F. and Willems, B. and Kalogera, V.
               and Rasio, F.~A.},
    title   = {Interacting Binaries with Eccentric Orbits. {II}.
               Secular Orbital Evolution Due to Non-Conservative
               Mass Transfer},
    journal = {\apj},
    year    = {2009},
    volume  = {702},
    pages   = {1387--1392},
    doi     = {10.1088/0004-637X/702/2/1387},
    eprint  = {0903.0621},
    archiveprefix = {arXiv}
}

@ARTICLE{Zahn1977,
       author = {{Zahn}, J.-P.},
        title = "{Tidal friction in close binary stars}",
      journal = {\aap},
         year = 1977,
       volume = {57},
        pages = {383-394},
       adsurl = {https://ui.adsabs.harvard.edu/abs/1977A&A....57..383Z},
}

@ARTICLE{ScholzStephens1987,
       author = {{Scholz}, F.~W. and {Stephens}, M.~A.},
        title = "{K-Sample Anderson-Darling Tests}",
      journal = {Journal of the American Statistical Association},
         year = 1987,
       volume = {82},
       number = {399},
        pages = {918-924},
          doi = {10.1080/01621459.1987.10478517},
}

@ARTICLE{Callingham2020,
       author = {{Callingham}, J.~R. and {Crowther}, P.~A. and {Williams}, P.~M. and {Tuthill}, P.~G. and {Han}, Y. and {Pope}, B.~J.~S. and {Marcote}, B.},
        title = "{Two Wolf-Rayet stars at the heart of colliding-wind binary Apep}",
      journal = {\mnras},
         year = 2020,
       volume = {495},
       number = {3},
        pages = {3323-3331},
          doi = {10.1093/mnras/staa1244},
       eprint = {2005.00542},
 primaryClass = {astro-ph.SR},
}

@ARTICLE{Smith2011,
       author = {{Smith}, Nathan},
        title = "{Explosions triggered by violent binary-star collisions: application to Eta Carinae and other eruptive transients}",
      journal = {\mnras},
         year = 2011,
       volume = {415},
       number = {3},
        pages = {2020-2024},
          doi = {10.1111/j.1365-2966.2011.18607.x},
       eprint = {1010.3770},
 primaryClass = {astro-ph.SR},
}

@ARTICLE{Wang2021,
       author = {{Wang}, Chen and {Podsiadlowski}, Philipp and {Han}, Zhanwen},
        title = "{Asymmetrical mass ejection from proto-white dwarfs and the formation of eccentric millisecond pulsar binaries}",
      journal = {\mnras},
         year = 2021,
       volume = {506},
       number = {3},
        pages = {4654-4666},
          doi = {10.1093/mnras/stab1926},
       eprint = {2101.12433},
 primaryClass = {astro-ph.SR},
}

@ARTICLE{BrandtPodsiadlowski1995,
       author = {{Brandt}, N. and {Podsiadlowski}, Ph.},
        title = "{The effects of high-velocity supernova kicks on the orbital properties and sky distributions of neutron-star binaries}",
      journal = {\mnras},
         year = 1995,
       volume = {274},
       number = {2},
        pages = {461-484},
          doi = {10.1093/mnras/274.2.461},
}

@ARTICLE{Crowther2010, 
       author = {{Crowther}, Paul A. and {Schnurr}, Olivier and {Hirschi}, Rapha{\"e}l and {Yusof}, Norhasliza and {Parker}, Richard J. and {Goodwin}, Simon P. and {Kassim}, Hasan Abu},
        title = "{The R136 star cluster hosts several stars whose individual masses greatly exceed the accepted 150 Msun stellar mass limit}",
      journal = {\mnras},
         year = 2010,
       volume = {408},
       number = {2},
        pages = {731-751},
          doi = {10.1111/j.1365-2966.2010.17167.x},
       eprint = {1007.3284},
 primaryClass = {astro-ph.SR},
}

@ARTICLE{HadjiDemetriou1969, 
       author = {{Hadjidemetriou}, J.~D.},
        title = "{Two-body problem with variable mass: A new approach}",
      journal = {Icarus},
         year = 1969,
       volume = {10},
        pages = {498-499},
          doi = {10.1016/0019-1035(69)90101-X},
}

@ARTICLE{HuangEtAl1956, 
       author = {{Huang}, Su-Shu},
        title = "{A Nova Theory of Close Binary Systems}",
      journal = {\aj},
         year = 1956,
       volume = {61},
        pages = {49},
          doi = {10.1086/107290},
}

@ARTICLE{Jeans1924,
       author = {{Jeans}, J.~H.},
        title = "{Cosmogonic problems associated with a secular decrease of mass}",
      journal = {\mnras},
         year = 1924,
       volume = {85},
        pages = {2-11},
          doi = {10.1093/mnras/85.1.2},
}

@ARTICLE{VinkdeKoterLamers2001,  
       author = {{Vink}, Jorick S. and {de Koter}, A. and {Lamers}, H.~J.~G.~L.~M.},
        title = "{Mass-loss predictions for O and B stars as a function of metallicity}",
      journal = {\aap},
         year = 2001,
       volume = {369},
        pages = {574-588},
          doi = {10.1051/0004-6361:20010127},
       eprint = {astro-ph/0101509},
 primaryClass = {astro-ph},
}

@ARTICLE{Williams2012,  
       author = {{Williams}, P.~M. and {van der Hucht}, K.~A. and others},
        title = "{Dust formation in the Wolf-Rayet system WR 48a}",
      journal = {\mnras},
         year = 2012,
       volume = {420},
        pages = {2526},
}

@ARTICLE{Fernandez-Martin2013,
  author  = {{Fern{\'a}ndez-Mart{\'i}n}, A. and {V{\'i}lchez}, J. M. and
             {P{\'e}rez-Montero}, E. and {Candian}, A. and {S{\'a}nchez}, S. F. and
             {Mart{\'i}n-Gord{\'o}n}, D. and {Riera}, A.},
  title   = {{Integral field spectroscopy of M1-67. A Wolf-Rayet nebula with
             luminous blue variable nebula appearance}},
  journal = {Astronomy \& Astrophysics},
  year    = {2013},
  volume  = {554},
  pages   = {A104},
  doi     = {10.1051/0004-6361/201220773},
}

@ARTICLE{Toala2015,
  author  = {{Toal{\'a}}, J. A. and {Guerrero}, M. A. and {Ramos-Larios}, G. and
             {Guzm{\'a}n}, V.},
  title   = {{WISE morphological study of Wolf-Rayet nebulae}},
  journal = {Astronomy \& Astrophysics},
  year    = {2015},
  volume  = {578},
  pages   = {A66},
  doi     = {10.1051/0004-6361/201525706},
}

@ARTICLE{Kozai1962,
  author  = {{Kozai}, Yoshihide},
  title   = {{Secular perturbations of asteroids with high inclination and
             eccentricity}},
  journal = {The Astronomical Journal},
  year    = {1962},
  volume  = {67},
  pages   = {591--598},
  doi     = {10.1086/108790},
}

@ARTICLE{Lidov1962,
  author  = {{Lidov}, M. L.},
  title   = {{The evolution of orbits of artificial satellites of planets under
             the action of gravitational perturbations of external bodies}},
  journal = {Planetary and Space Science},
  year    = {1962},
  volume  = {9},
  pages   = {719--759},
  doi     = {10.1016/0032-0633(62)90129-0},
}

@ARTICLE{Naoz2016,
  author  = {{Naoz}, Smadar},
  title   = {{The Eccentric Kozai-Lidov Effect and Its Applications}},
  journal = {Annual Review of Astronomy and Astrophysics},
  year    = {2016},
  volume  = {54},
  pages   = {441--489},
  doi     = {10.1146/annurev-astro-081915-023315},
}

@ARTICLE{Dsilva2020,
  author  = {{Dsilva}, K. and {Shenar}, T. and {Sana}, H. and {Marchant}, P.},
  title   = {{A spectroscopic multiplicity survey of Galactic Wolf-Rayet stars.
             I. The northern WC sequence}},
  journal = {Astronomy \& Astrophysics},
  year    = {2020},
  volume  = {641},
  pages   = {A26},
  doi     = {10.1051/0004-6361/202038446},
}

@ARTICLE{Dsilva2022,
  author  = {{Dsilva}, K. and {Shenar}, T. and {Sana}, H. and {Marchant}, P.},
  title   = {{A spectroscopic multiplicity survey of Galactic Wolf-Rayet stars.
             II. The northern WNE sequence}},
  journal = {Astronomy \& Astrophysics},
  year    = {2022},
  volume  = {664},
  pages   = {A93},
  doi     = {10.1051/0004-6361/202142729},
}

@ARTICLE{Dsilva2023,
  author  = {{Dsilva}, K. and {Shenar}, T. and {Sana}, H. and {Marchant}, P.},
  title   = {{A spectroscopic multiplicity survey of Galactic Wolf-Rayet stars.
             III. The northern late-type nitrogen-rich sample}},
  journal = {Astronomy \& Astrophysics},
  year    = {2023},
  volume  = {674},
  pages   = {A88},
  doi     = {10.1051/0004-6361/202244308},
}

@ARTICLE{SmithConti2008,
  author  = {{Smith}, Nathan and {Conti}, Peter S.},
  title   = {{On the Role of the WNH Phase in the Evolution of Very Massive Stars:
             Enabling the LBV Instability with Feedback}},
  journal = {The Astrophysical Journal},
  year    = {2008},
  volume  = {679},
  number  = {2},
  pages   = {1467--1477},
  doi     = {10.1086/586885},
}

@ARTICLE{han2020apep,
  author  = {Han, Yinuo and Tuthill, Peter G. and Lau, Ryan M. and Soulain, Antoine and
             Callingham, Joseph R. and Williams, Peter M. and Crowther, Paul A. and
             Pope, Benjamin J. S. and Marcote, Benito},
  title   = {{The extreme colliding-wind system Apep: resolved imagery of the central
             binary and dust plume in the infrared}},
  journal = {MNRAS},
  year    = {2020},
  volume  = {498},
  number  = {4},
  pages   = {5604--5619},
  doi     = {10.1093/mnras/staa2349},
}

@ARTICLE{schweickhardtetal99,
       author = {{Schweickhardt}, J. and {Schmutz}, W. and {Stahl}, O. and {Szeifert}, Th. and {Wolf}, B.},
        title = "{Revised mass determination of the super massive Wolf-Rayet star WR 22}",
      journal = {\aap},
         year = 1999,
       volume = {347},
        pages = {127-136},
}

@ARTICLE{gosset-nase16,
       author = {{Gosset}, E. and {Naz{\'e}}, Y.},
        title = "{The X-ray light curve of the massive colliding wind Wolf-Rayet + O binary WR 21a}",
      journal = {\aap},
         year = 2016,
       volume = {590},
          eid = {A113},
        pages = {A113},
          doi = {10.1051/0004-6361/201527427},
       eprint = {1604.05005},
 primaryClass = {astro-ph.SR},
}

@ARTICLE{frohlich-regaly-25,
       author = {{Fr{\"o}hlich}, Vikt{\'o}ria and {Reg{\'a}ly}, Zsolt},
        title = "{Life in the dark: Potential urability of the moons of rogue planets}",
      journal = {\aap},
         year = 2025,
        month = dec,
       volume = {704},
          eid = {A349},
        pages = {A349},
          doi = {10.1051/0004-6361/202556673},
archivePrefix = {arXiv},
       eprint = {2511.03392},
 primaryClass = {astro-ph.EP},
       adsurl = {https://ui.adsabs.harvard.edu/abs/2025A&A...704A.349F}
}

@ARTICLE{regaly-etal-22,
       author = {{Reg{\'a}ly}, Zsolt and {Fr{\"o}hlich}, Vikt{\'o}ria and {Vink{\'o}}, J{\'o}zsef},
        title = "{Lost in Space: Companions' Fatal Dance around Massive Dying Stars}",
      journal = {\apj},
         year = 2022,
        month = dec,
       volume = {941},
       number = {2},
          eid = {121},
        pages = {121},
          doi = {10.3847/1538-4357/aca1ba},
archivePrefix = {arXiv},
       eprint = {2211.04600},
 primaryClass = {astro-ph.SR},
       adsurl = {https://ui.adsabs.harvard.edu/abs/2022ApJ...941..121R}
}

@ARTICLE{frohlich-etal-23,
       author = {{Fr{\"o}hlich}, Vikt{\'o}ria and {Reg{\'a}ly}, Zsolt and {Vink{\'o}}, J{\'o}zsef},
        title = "{Double neutron star formation via consecutive type II supernova explosions}",
      journal = {\mnras},
         year = 2023,
        month = aug,
       volume = {523},
       number = {4},
        pages = {4957-4969},
          doi = {10.1093/mnras/stad1788},
archivePrefix = {arXiv},
       eprint = {2306.07099},
 primaryClass = {astro-ph.SR},
       adsurl = {https://ui.adsabs.harvard.edu/abs/2023MNRAS.523.4957F}
}

@ARTICLE{2020SciPy-NMeth,
  author  = {Virtanen, Pauli and Gommers, Ralf and Oliphant, Travis E. and
             Haberland, Matt and Reddy, Tyler and Cournapeau, David and
             Burovski, Evgeni and Peterson, Pearu and Weckesser, Warren and
             Bright, Jonathan and {van der Walt}, St{\'e}fan J. and
             Brett, Matthew and Wilson, Joshua and Millman, K. Jarrod and
             Mayorov, Nikolay and Nelson, Andrew R. J. and Jones, Eric and
             Kern, Robert and Larson, Eric and Carey, C J and
             Polat, {\.I}lhan and Feng, Yu and Moore, Eric W. and
             {VanderPlas}, Jake and Laxalde, Denis and Perktold, Josef and
             Cimrman, Robert and Henriksen, Ian and Quintero, E. A. and
             Harris, Charles R. and Archibald, Anne M. and
             Ribeiro, Ant{\^o}nio H. and Pedregosa, Fabian and
             {van Mulbregt}, Paul and {SciPy 1.0 Contributors}},
  title   = {{SciPy 1.0: Fundamental Algorithms for Scientific Computing
             in Python}},
  journal = {Nature Methods},
  year    = {2020},
  volume  = {17},
  pages   = {261--272},
  doi     = {10.1038/s41592-019-0686-2},
}

@BOOK{HairerNorsettWanner1993,
  author    = {Hairer, Ernst and N{\o}rsett, Syvert P. and Wanner, Gerhard},
  title     = {Solving Ordinary Differential Equations I: Nonstiff Problems},
  edition   = {2nd},
  series    = {Springer Series in Computational Mathematics},
  volume    = {8},
  publisher = {Springer-Verlag},
  address   = {Berlin, Heidelberg},
  year      = {1993},
  doi       = {10.1007/978-3-540-78862-1},
}

@ARTICLE{Lau2020,
  author  = {Lau, R. M. and Hankins, M. J. and Han, Y. and Endo, I. and
             Hummel, C. A. and Ressler, M. E. and Sanchez-Bermudez, J. and
             Smith, N. and Stee, P. and Tanaka, D. and Moffat, A. F. J. and
             Onaka, T. and Williams, P. M.},
  title   = {{Resolving Decades of Periodic Spirals from the Wolf-Rayet
             Dust Factory WR 112}},
  journal = {The Astrophysical Journal},
  year    = {2020},
  volume  = {900},
  number  = {2},
  pages   = {190},
  doi     = {10.3847/1538-4357/abaab8},
}
\bibliographystyle{aasjournal}



\end{document}